\documentclass[preprint,showpacs,showkeys,preprintnumbers,aps,prd,amsfonts,eqsecnum,nofootinbib,
floatfix]
{revtex4-1}
\usepackage{graphicx,epsfig}
\usepackage[dvips]{color}
\usepackage{soul}
\def\gsim{\lower0.5ex\hbox{$\:\buildrel >\over\sim\:$}}
\def\lsim{\lower0.5ex\hbox{$\:\buildrel <\over\sim\:$}}

\usepackage{color}
\usepackage{multirow}
\usepackage{slashed}

\begin{document}
\preprint{CUMQ/HEP 163}

\title{\large Top Quark Pair Production and  Asymmetry at the Tevatron and LHC in Left-Right Models}

\author{Mariana Frank$^{a}$}\email{mfrank@alcor.concordia.ca}
\author{Alper Hayreter$^{a}$}\email{alper@physics.concordia.ca}
\author{Ismail Turan$^{b}$}\email{ituran@physics.carleton.ca}
\affiliation{$^{a}$ Department of Physics, Concordia University, 7141 Sherbrooke St. 
West, Montreal, Quebec, CANADA H4B 1R6\\
 $^{b}$ Ottawa-Carleton 
Institute of Physics,
Carleton University,
1125 Colonel By Drive
Ottawa, Ontario, Canada, K1S 5B6}

\date{\today}
             
\begin{abstract}
In light of the recent  measurements of the top quark forward-backward asymmetry at the Fermilab Tevatron experiment, which
in some regions of the parameter space shows a discrepancy of 3$\sigma$ compared to the SM prediction, we analyze top quark
pair production and asymmetry in the context of left-right models both  at the Tevatron and LHC. We use the minimal manifest
left-right model and an asymmetric left-right model where gauge couplings and flavor mixing in the right-handed sector are
allowed to differ from those in the left-handed sector. We explore the consequences of including effects from $W_R$ and
$Z_R$ gauge bosons, consistent with phenomenological constraints from meson mixing and new bounds from ATLAS and CMS, for
the $t \bar{t}$ cross section, invariant mass distribution and forward-backward asymmetry at the Tevatron, and predict their
values at the LHC. We show that, choosing parameter benchmarks for the model while preserving agreement with collider,
electroweak precision, and flavor violation data, the generic left-right model cannot account for the large deviations of the
observed asymmetry at the Tevatron and also that it predicts very small charge asymmetries at the LHC. 

\end{abstract}


\maketitle

\section{Introduction}

Measurements of top production and decays are of particular interest for particle theorists as they likely will shed light 
on the mechanism of electroweak symmetry breaking. The Tevatron has produced such measurements, and more are expected to 
come from the LHC. For instance, the $t {\bar t}$ total cross section, as well as the differential cross section with 
respect to the $t {\bar t}$ invariant mass, both of which are sensitive to a variety of beyond the standard model (BSM)
scenarios of particles decaying into $t {\bar t}$ pairs, are completely consistent with the standard model (SM)
\cite{cdf-ttbar,Aaltonen:2009iz,D0:dsdmtt}.

But recently both the CDF and D0 collaborations have measured the forward-backward asymmetry of the top quark pairs,
$A_{FB}^{t\bar{t}}$ \cite{cdf,newcdf,d0}. Based on a data sample of $5.3$~fb$^{-1}$~\cite{newcdf}, the asymmetries,
evolved to the parton level\footnote{Here and throughout the paper, parton-level is used in the same meaning described in
\cite{newcdf}. It refers to deconvolving from the data, like detector efficiencies, jet algorithm, selection
efficiencies, background etc. See \cite{newcdf} for more details.}, are:
\begin{eqnarray}
A^{t \bar t}(|\Delta_y| < 1)  &=& 0.026 \pm 0.118
\nonumber\\
A^{t \bar t}(|\Delta_y| \ge 1)  &=& 0.611 \pm 0.256 
\nonumber\\
A^{t \bar t}(M_{t \bar t}<450~{\rm GeV})  &=& -0.116 \pm 0.153 \nonumber\\
A^{t \bar t}(M_{t \bar t} \ge 450~{\rm GeV}) &=& 0.475 \pm 0.114
\label{eq:newcdf}
\end{eqnarray}
in the $t \bar{t}$ rest frame (with $m_t = 175$ GeV). In the SM the asymmetry is  produced mainly through one-loop QCD
corrections, with a smaller contribution from electroweak $t \bar{t}$ production, and is stable with respect to corrections
from QCD threshold resummation~\cite{Almeida:2008ug}. The next-to-leading-order (NLO) SM predictions  at parton level
are, by comparison 
~\cite{Antunano:2007da,Bowen:2005ap,mynlo}
\begin{eqnarray}
A_{\rm SM}^{t \bar t}(|\Delta_y| < 1)  &=& 0.039 \pm 0.006
\nonumber\\
A_{\rm SM}^{t \bar t}(|\Delta_y| \ge 1  &=& 0.123 \pm 0.008 
\nonumber\\
A_{\rm SM}^{t \bar t}(M_{t \bar t}<450~{\rm GeV})  &=& 0.040 \pm 0.006
 \nonumber\\
A_{\rm SM}^{t \bar t}(M_{t \bar t} \ge 450~{\rm GeV}) &=& 0.088 \pm 0.013
\label{eq:SMnlo}
\end{eqnarray}

We note that there has been a recent calculation of the asymmetry including electroweak corrections to ${\cal O}(\alpha^2)$
terms, as well as interferences with the QCD diagrams \cite{Hollik:2011ps}. It seems that SM asymmetry receives
non-negligible same-sign contributions from the electroweak sector so that, except the region with $M_{t \bar t}>450$ GeV,
the observed deviation between theory and experiment diminishes. 

As the deviation from the expected and the measured asymmetry is large, this has been interpreted as a signal for new 
physics (NP), in particular a signal for a below-TeV scale physics. A large  variety of models has been employed to resolve
the discrepancy. These models invoke new particles and new interactions to explain the discrepancy. In general, one can
classify these models according to the new mediators of the new physics as $(1)$ $t-$channel boson mediators (scalars or
vectors, such as $W^\prime$ or $Z^\prime$) with  flavor-violating couplings to right-handed up quarks \cite{Jung:2009jz}, 
$(2)$ $s-$channel mediators, color sextet or color anti-triplet scalar particles coupling with flavor-violating couplings to
up and top quarks, such as  \cite{Sehgal:1987wi}, or $(3)$ new flavor multiplets coupling to quarks in a
flavor-symmetric way \cite{Ligeti:2011vt}. Comparative studies of various models also exist, and it was shown that
$s-$channel particles used to explain the anomaly have maximal axial couplings, while $t-$channel particles exhibit maximal
flavor violating couplings \cite{Jung:2009pi}. As well, a number of analysis have appeared, which study the implications of
models which predict large asymmetry for LHC phenomenology \cite{Hewett:2011wz}. These models have been studied individually,
or in a group, to extract some global features which would insure generating a large asymmetry while contributing a
negligible amount to the cross section, and to classify general features. A recent analysis \cite{Blum:2011fa} concludes
that, among scalar mediated-processes, only the $t-$channel exchange of a QCD-singlet, weak doublet scalar is consistent 
with flavor and electroweak  constraints,  and does not conflict with the collider data obtained so far. 
  
Although these models have been shown to produce a large asymmetry, they all appear designed specifically to resolve this
problem, are sometimes insufficiently justified, and thus they seem disconnected from other low energy phenomenology
constraints. In all models, large flavor-violation in the $t-u$ or $t-d$ quark sectors is enhanced, while flavor-changing in
the other sectors is suppressed. The question remains of whether such asymmetry can be obtained by employing a known and
well-studied NP model. In particular, what is the prediction of such a model (allowing for maximum flexibility) and how
important is  for the prediction of the asymmetry to impose  the requirement that the model satisfies known phenomenological
constraints. We propose to investigate here the effect on the asymmetry and $t{\bar t}$ production cross section emerging
from $W_R$ and $Z_R$ bosons in the left-right symmetric model. This model satisfies some definite conditions: 

\begin{itemize}
\item It is one of the simplest and most natural extensions of the SM;
\item  It contains additional particles in both the $s-$ and $t-$channels which could enhance the forward-backward asymmetry,
but also the $ t {\bar t}$ cross section; 
\item It has been thoroughly investigated and constrained through many analysis, and in particular CDF and D0 have put 
limits on extra boson masses;  
\item  More information and testing of the model will be provided soon by LHC (some recent bounds from colliders are
discussed later).
\end{itemize}

We first perform an analysis of the $t {\bar t}$ pair production and forward-backward asymmetry at the Tevatron, then we
explore the signal at LHC, for both the cross section and possible asymmetries testable at the LHC. As we wish to allow the
model to be as general as possible, we rely on a generic model, without constraining masses, mixing parameters or gauge
couplings, but impose constraints coming from low energy phenomenology, mainly $K$ and $B$ physics, but also collider
restrictions coming from the Tevatron. As the LHC data would be available fast, and  the constraints on particular models 
are rapidly changing, we are  motivated by the fact that the LHC collaborations are now analyzing unprecedented amounts of
top data that will clearly rule out models. Thus a clear expectation of model predictions for the LHC is timely.

We will work in a parametrization in which the quark mixing matrices in the left- and right-handed sectors ($V_{CKM}^L$ and
$V_{CKM}^R$) are allowed to differ, and so do the $SU(2)_{L,R}$ coupling constants $g_L$ and $g_R$. We discuss the case in
which $V_{CKM}^R=V_{CKM}^L$ and $g_R=g_L$ as a particular case of a larger family of solutions. We are interested in the
asymmetries and cross sections which can be obtained in left-right models which satisfy the low energy constraints, but also
investigate values of the  asymmetries in the model when we relax the known constraints. We perform the same analysis for 
the LHC, where we investigate the cross section and LHC asymmetries at both $\sqrt{s}=7$ TeV and $\sqrt{s}=14$ TeV. 
Prospects for differentiating the left-right symmetric model from other BSM scenarios are outlined.

Our paper is organized as follows: in Section \ref{sec:lrm} we describe briefly our model. We proceed to evaluate the
top-pair production and forward-backward asymmetry at the Tevatron in Section \ref{sec:Tevatron}. Section \ref{sec:LHC} we
explore the model predictions for the LHC. We discuss our findings and conclude in Section \ref{sec:conclusion}.  

\section{Left-Right Symmetric Models}
\label{sec:lrm}

We assume a generic left-right (LR) symmetric model based on the gauge group $SU(3)_C \times SU(2)_L \times SU(2)_R \times
U(1)_{B-L}$ \cite{Pati:1974yy}. The matter fields of this model consist of three families of quark and lepton fields with 
the following transformations under the gauge group:
\begin{eqnarray}
Q_L^i &=&\left( \begin{array}{c} u_L^i \\ d_L^i \end{array} \right)  \sim \left ( 3, 2, 1, 1/3 \right ),~~
Q_R^i =\left( \begin{array}{c} u_R^i \\ d_R^i \end{array} \right) \sim \left ( 3,1, 2, 1/3 \right ),\nonumber \\
L_L^i& = &\left( \begin{array}{c} \nu_L^i \\ e_L^i \end{array} \right) \sim\left( 1,2, 1, -1 \right),~~
L_R^i= \left( \begin{array}{c} \nu_R^i \\ e_R^i \end{array} \right) \sim \left ( 1,1, 2, -1 \right ),
\end{eqnarray}
with the numbers in the brackets representing the quantum numbers under, respectively $SU(3)_c ,~ SU(2)_L , ~SU(2)_R$ and
$U(1)_{B-L}$. The gauge bosons of the left right model are $\gamma, ~Z_L,~Z_R$, in the neutral sector, and $W_L^\pm,
~W_R^\pm$ in the charged one. The Higgs sector necessary to break the left-right model  consists of one bidoublet:
\begin{eqnarray}
\displaystyle
\Phi&&= \left( \begin{array}{cc} \phi^0_{1} &\phi^+_{2} \\ \phi^-_{1} &
 \phi^0_{2} \end{array} \right) \sim \left (1,2,2,0 \right),
\end{eqnarray}
with the Vacuum Expectation Values (VEVs)
\begin{eqnarray}
\displaystyle
\langle\Phi\rangle&&= \left( \begin{array}{cc} k &0 \\ 0 & k^\prime\end{array} \right) .
\end{eqnarray}
Additional Higgs multiplets are needed to break the symmetry to $SU(2)_L \times U(1)_Y$ and to generate a large mass of $W_R$
relative to $W_L$. Higgs triplets are a popular choice, as their VEV  can also produce a large $M_{W_R}$ mass and generate a
large Majorana neutrino mass through the seesaw mechanism\cite{Mohapatra:1979ia}
\begin{eqnarray}
\Delta_{L} = \left(\begin{array}{cc}
\frac {\Delta_L^+}{\sqrt{2}}&\Delta_L^{++}\\
\Delta_{L}^{0}&-\frac{\Delta_L^+}{\sqrt{2}}
\end{array}\right) \sim (1,3,1,2),~~~
\Delta_{R}  =
\left(\begin{array}{cc}
\frac {\Delta_R^+}{\sqrt{2}}&\Delta_R^{++}\\
\Delta_{R}^{0}&-\frac{\Delta_R^+}{\sqrt{2}}
\end{array}\right) \sim (1,1,3,2).
\end{eqnarray}
with VEVs, $v_{\Delta_{L}} \equiv v_L= 0$ and $v_{\Delta_{R}} \equiv v_R $. 

The charged gauge bosons $W_L$ and $W_R$ mix to form mass eigenstates $W_1$ and $W_2$
\begin{eqnarray}
W_L&=&W_1\cos \xi -W_2 \sin \xi 
\nonumber \\
W_R&=&e^{i\omega}(W_1\sin \xi +W_2\cos \xi)
\end{eqnarray}
with $\xi$ a mixing angle and $\omega$ a CP violating phase.  If $\xi$ is small, then $W_L$ and $W_R$ approximately coincide
with $W_1$ and $W_2$. The mass matrix for the charged bosons is
\begin{eqnarray}
M^2_{W}  &=&\frac{1}{4}
\left(\begin{array}{cc}
g_L^2(|k|^2 +|k^{\prime}|^2+2|v_L|^2) & -2g_Lg_R k^\prime k^\star\\
-2g_Lg_Rk^{\prime\star} k & g_R^2(|k|^2 +|k^{\prime}|^2+2|v_R|^2)
\end{array}\right)
\end{eqnarray}
For $|v_R| \gg (|k|, |k^\prime|)\gg |v_L|$ the masses become approximately
\begin{eqnarray}
M_1^2&\simeq&\frac14g_L^2(|k|^2 +|k^{\prime}|^2), \nonumber\\
M_2^2&\simeq&\frac12g_R^2|v_R|^2
\end{eqnarray}
and the mixing angle is
\begin{equation}
\xi \simeq \pm\frac{g_L}{g_R}\frac{2|k k^\prime|}{|v_R|^2}\ .
\end{equation}

The right-handed bosons contribute to the charged and neutral currents for the quarks, which is
\begin{eqnarray}
{\cal L}_{CC}&=&\frac{g_L}{\sqrt{2}}{\bar u}{_{iL}\gamma _\mu V^L_{CKM\,ij}d_{jL} W^{\mu +}_L +\frac{g_R}{\sqrt{2}}{\bar
u}_{iR}}\gamma _\mu V^R_{CKM\,ij}d_{jR} W^{\mu +}_R\;, \\
{\cal L}_{NC}&=&\frac{g_L}{\cos \theta_W} \left [ {\bar u}_{i} \gamma _\mu \left (T^u_{3} P_L-e_u \sin^2 \theta_W \right)
u_{j}+ {\bar d}_{i} \gamma _\mu \left (T_3^d P_L -e_d \sin^2 \theta_W \right)  d_{j}\right] Z^{\mu }_L\nonumber \\
& +& {g_R \cos \phi}\left [ {\bar u}_{i} \gamma _\mu \left (T^u_{3} P_R-\frac16 \tan^2 \phi \right) u_{j}+ {\bar d}_{i}
\gamma _\mu \left (T^d_{3} P_R-\frac16 \tan^2 \phi \right) d_{j} \right] Z^{\mu}_R \nonumber\\
\end{eqnarray}
where $P_{L,R}=(1\mp\gamma_5)/2$ and $\displaystyle \sin \phi= \frac{g_{B-L}}{\sqrt {g_{B-L}^2+g_R^2}} ~(\sin\phi=
\tan\theta_W~{\rm for}~g_R=g_L) $ and similarly for the leptons, which are allowed to mix with different CKM-type matrices.
We adopt the Wolfenstein parametrization for the CKM matrix $V^L_{CKM}$\cite{Nakamura:2010zzi}
\begin{eqnarray}
V_{CKM}^L  =
\left(\begin{array}{ccc}
1-\frac{\lambda^2}{2}&\lambda&A\lambda^3 (\rho-i\eta)\\
-\lambda&1-\frac{\lambda^2}{2}&a\lambda^2\\
A\lambda^3(1-\rho-i\eta)&-A\lambda^2&1
\end{array}\right)~.
\end{eqnarray}

For the right-handed CKM matrix, there are several left-right scenarios which appear in the literature:
\begin{itemize}
\item{ In {\it manifest LR symmetric models} \cite{Senjanovic:1978ev}, the CP violation is generated by complex Yukawa
couplings, while the vevs of the Higgs fields remain real. This implies the same mixing for right and left-handed quarks,
$V_{CKM}^R= V_{CKM}^L$, where $V_{CKM}^L$ is the usual Cabibbo-Kobayashi-Maskawa matrix, and equal gauge couplings for
$SU(2)_L$ and $SU(2)_R$, $g_R=g_L$.}

\item{ In {\it pseudo-manifest LR symmetry}, both CP and P symmetries are spontaneously broken \cite{Harari:1983gq}, such
that the Yukawa couplings are real. In this case the left and right handed quark mixings are related through $V_{CKM}^R=
V_{CKM}^{L \star}K$, with $K$ a diagonal phase matrix. Here as well, $g_R=g_L$. }

\item{In {\it asymmetric LR symmetry}, left-right symmetry is assumed to be fundamental, superseding the Higgs, Yukawa, or
fermion structure \cite{Langacker:1989xa}. Here  arbitrary mixing between the second and third generations, or between the
first and third generations are allowed (within unitarity constraints). To simplify the notation, we drop the CKM subscript
and, following \cite{Langacker:1989xa}, denote the parametrizations as $(A)$ and $(B)$, where
\begin{eqnarray}
\label{scenAB}
V^R_{(A)} = \left(\begin{array}{ccc}
1&0&0\\
0&\cos \alpha&\pm \sin\alpha\\
0& \sin \alpha& \mp \cos \alpha
\end{array}\right) ,~~~
V_{(B)}^R  = \left(\begin{array}{ccc}
0&1&0\\
\cos\alpha &0&\pm \sin \alpha\\
\sin \alpha&0& \mp \cos \alpha
\end{array}\right),
\end{eqnarray}
with $\alpha$ an arbitrary angle $(-\pi/2 \le \alpha \le \pi/2)$. In parametrization $(A)$, depending on the values of
$\alpha$, the dominant coupling could be $V^R_{ts}$ while in $(B)$, the dominant coupling could be $V^R_{td}$. The $(A)$ and
$(B)$ parametrizations are chosen to  allow relaxing the mass limit on $W_R$ while obeying the restrictions on $\Delta m_K$ 
without fine-tuning.}
\end{itemize}

The form of the CKM matrix in the right-handed quark sector affects low energy phenomenology, in particular processes with
flavor violation, and thus restricts the mass $M_{W_R}$ and the mixing angle $\xi$. These have been analyzed recently in
\cite{Frank:2010qv,Frank:2010cj}. (For an alternative analysis, concentrating on the CP violation properties of the model,
see also \cite{Kiers:2002cz}.)
  
The constraints on the parameter space of the left-right model, mostly from flavor violating processes, which are relevant to
the study of $W_R$ phenomenology, come from $K^0-{\bar K}^0$ mixing, $B_d^0-{\bar B}_d^0$ and $B_s^0-{\bar B}_s^0$ mixing,
and $b \to s \gamma$. These constraints depend on several parameters and are difficult to summarize analytically; however,
they are included in the evaluation of the $t {\bar t}$ cross section and forward-backward asymmetry, analyzed in the next
section. 

We also include restrictions imposed by the available data from ATLAS which seems to rule out a $Z^\prime$ resonance with
$M_{Z^\prime} <950$ GeV, with the exact limit depending on specific models and specific assumptions \cite{Aad:2011xp}. A
recent talk at the European Physics Society meeting \cite{atlas} reports new bounds on $Z^\prime$ mass, with 50 times more
data ($\sim 2\, {\rm fb^{-1}}$) and with new bounds varying from $1.5$ TeV to $1.8$ TeV depending on the models. Similarly
there are new bounds from the CMS and D0 collaborations \cite{cms,Abazov:2010ti} with total integrated luminosity $1.1\, {\rm
fb^{-1}}$ and $5.4\, {\rm fb^{-1}}$, respectively. While the bounds from CMS are very similar to the ones from ATLAS, D0
bounds are somewhat weaker. A relevant study by Nemevsek et al \cite{Nemevsek:2011hz} on the bound on $W_R$ mass using the 33
pb$^{-1}$ LHC data at 7 TeV reports $M_{W_R}>1.4$ TeV, but is also spectrum specific and depends on whether the right-handed
neutrino is Majorana or Dirac and whether it is lighter or heavier than $M_{W_R}$. We assume the right-handed neutrino
heavier than $M_{W_R}$ so that the above bound is evaded.

For the evaluation of the cross section and the asymmetry, we have chosen two benchmark parameter sets for each of Model A,
Model B and Manifest left-right symmetric models, defined as previously. To select particular benchmark points, we used
the results of our previous parameter scans over $M_{W_R},\, \sin \alpha, M_{H^\pm}$ and $g_R/g_L$ in
\cite{Frank:2010qv,Frank:2010cj} where we have presented restrictions over the parameter space obtained by imposing low
energy constraints from meson mixings and  $b \to s \gamma$ branching ratio, as well as collider constraints on production of
extra gauge bosons. The parameter scan leaves very small allowed regions where the $W_R$ is light, and/or the flavor
violation from the right-handed sector is significant. These points  were  explicitly chosen  from all the allowed
parameter space to maximize flavor-violation in the right-handed quark sector, for both light and heavy $M_{W_R}$ scenarios.
 While we work with these choices, we shall comment on the effect of varying the chosen sets in the parameter space. The
parameter sets for each model, namely Set I and Set II, that are used in our calculations in accordance with those
constraints are given in Table \ref{params}.  We include, in addition to the Set I and Set II, a left-right scenario for each
of the three models which is not subjected to experimental constraints  as in  \cite{Frank:2010qv,Frank:2010cj}, which
we call the {\it Unconstrained LR Set}. We require that this model is roughly consistent with collider limits on the $t{\bar
t}$ cross section. Our aim is to show the effects of experimental restrictions on the parameter space and highlight that
``relaxing" them  can produce large asymmetries.  

\begin{table}[htb]
\begin{tabular*}{0.98\textwidth}{@{\extracolsep{\fill}} cccccccccc} 
\hline\hline
& \multicolumn{3}{c}{Manifest} & \multicolumn{3}{c}{Model A} & \multicolumn{3}{c}{Model B} \\ \cline{2-10}
& ~Set I~ & ~Set II~ & Uncons. & ~Set I~ & ~Set II~ & Uncons. & ~Set I~ & ~Set II~ & Uncons. \\ 
\multicolumn{1}{c}{~~$M_{W_R}$(GeV)~~} & 700  & 1500 & 500 & 700  & 1000 & 500 & 1100 & 1300 & 500 \\ 
\multicolumn{1}{c}{~~$M_{Z_R}$(GeV)~~} & 1172 & 2511 & 837 & 2189 & 1674 & 734 & 3441 & 2176 & 734 \\ 
\multicolumn{1}{c}{~~$g_R/g_L$~~}      & 1    & 1    & 1   & 0.6  & 1    & 2   & 0.6  & 1    & 2   \\ 
\multicolumn{1}{c}{~~$\sin\alpha$~~}   & -    & -    & -   & 0.5  & 0.25 & 0.7 & -0.2 & -0.1 & 0.7 \\ \hline\hline    
\end{tabular*}
\caption{Benchmark points Set I, Set II and Unconstrained for left-right symmetric models: Manifest,  Model A, and Model B, 
used throughout the analysis. Note that $M_{Z_R}$ is fixed when a value for $M_{W_R}$ is chosen but $M_{Z_R}$ values are
included for reference.}
\label{params}
\end{table}


\section{$t {\bar t}$  Cross section and forward-backward asymmetry at the Tevatron}
\label{sec:Tevatron}

The top quark pair production in $p\bar{p}$ collisions is mostly accomplished through $s$-channel quark-antiquark
annihilation (about 90\%) and much less so through $gg$ and $qg$ processes. The latest CDF and D0 measurements of the cross
section \cite{Aaltonen:2009iz} agree with the SM at the next-to-next-to-leading order (NNLO) prediction  
\cite{Cacciari:2008zb},
\begin{eqnarray}
\sigma^{\text{CDF II}}_{(p\bar{p}\rightarrow t\bar{t})} &=& 7.50 \pm 0.48 \,\text{pb}, \\ 
\sigma^{\text{NNLO}}_{(p\bar{p}\rightarrow t\bar{t})} &=& 7.39 \pm 0.55 \,\text{pb}.
\end{eqnarray}
We proceed to analyze the top-pair cross sections in the left-right models. For consistency, we evaluate here the cross
section in the SM, as well as in the LR models under scrutiny: the Manifest model, Model A and Model B,  for the Set I and
Set II for each model and, by comparison, for the Unconstrained set. Any new model must predict a cross section which agrees
with the experimental data, as the cross section is particularly sensitive to $s-$channel exotic resonances, thus restricting
the mass of the $Z_R$ boson in LR models.

In the calculation of $t\bar{t}$ production cross-sections we proceed as follows. We first calculate  the LO cross-sections
at $\sqrt{s}=1.96$ TeV with $m_t=172.5$ GeV, using {\tt CTEQ6M} parton distribution function (PDF) set to go from parton to 
$p {\bar p}$ cross sections. We then calculate  the NNLO cross section by multiplying the LO result with the $K$ factor
($K=1.3$ for Tevatron \cite{Cacciari:2008zb}) as in the SM. We assume for simplicity that the $K$-factors are universal, so
that the NP/SM ratios at LO and NNLO are the same, minimizing the impact of the NNLO corrections to the LR model
contributions (See our comments in the next paragraph).  We list the cross sections obtained in Table \ref{tbl:prodxsec-tev}.


\begin{table}[htb]
\begin{tabular*}{0.98\textwidth}{@{\extracolsep{\fill}}cccc} 
\hline\hline
{\bf SM} & & & \\
$\sigma_{\rm NNLO}$(pb) & $7.36\pm 0.007$ & & \\ 
{\bf Manifest}& ~{\bf Set I}~ & ~{\bf Set II}~ & {\bf Uncons.} \\ 
$\sigma_{\rm NNLO}$(pb) & $7.37\pm 0.007$ & $7.37\pm 0.007$ & $7.43\pm 0.008$ \\ 
{\bf Model A} & ~{\bf Set I}~ & ~{\bf Set II}~ & {\bf Uncons.} \\ 
$\sigma_{\rm NNLO}$(pb) & $7.36\pm 0.007$ & $7.37\pm 0.007$ & $8.35\pm 0.008$ \\ 
{\bf Model B} & ~{\bf Set I}~ & ~{\bf Set II}~ & {\bf Uncons.} \\ 
$\sigma_{\rm NNLO}$(pb) & $7.36\pm 0.007$ & $7.36\pm 0.007$ & $8.17\pm 0.008$ \\ \hline \hline   
\end{tabular*}
\caption{The NNLO $t\bar{t}$ production cross-sections at Tevatron  ($\sqrt{s}=1.96$ TeV) for the
SM, and Left-Right models: Manifest, Model A and Model B, for the benchmark points chosen.}
\label{tbl:prodxsec-tev}
\end{table}

\begin{figure}[htb] 
\begin{center}\hspace*{-1cm}
\includegraphics[width=3.0in,height=3.0in]{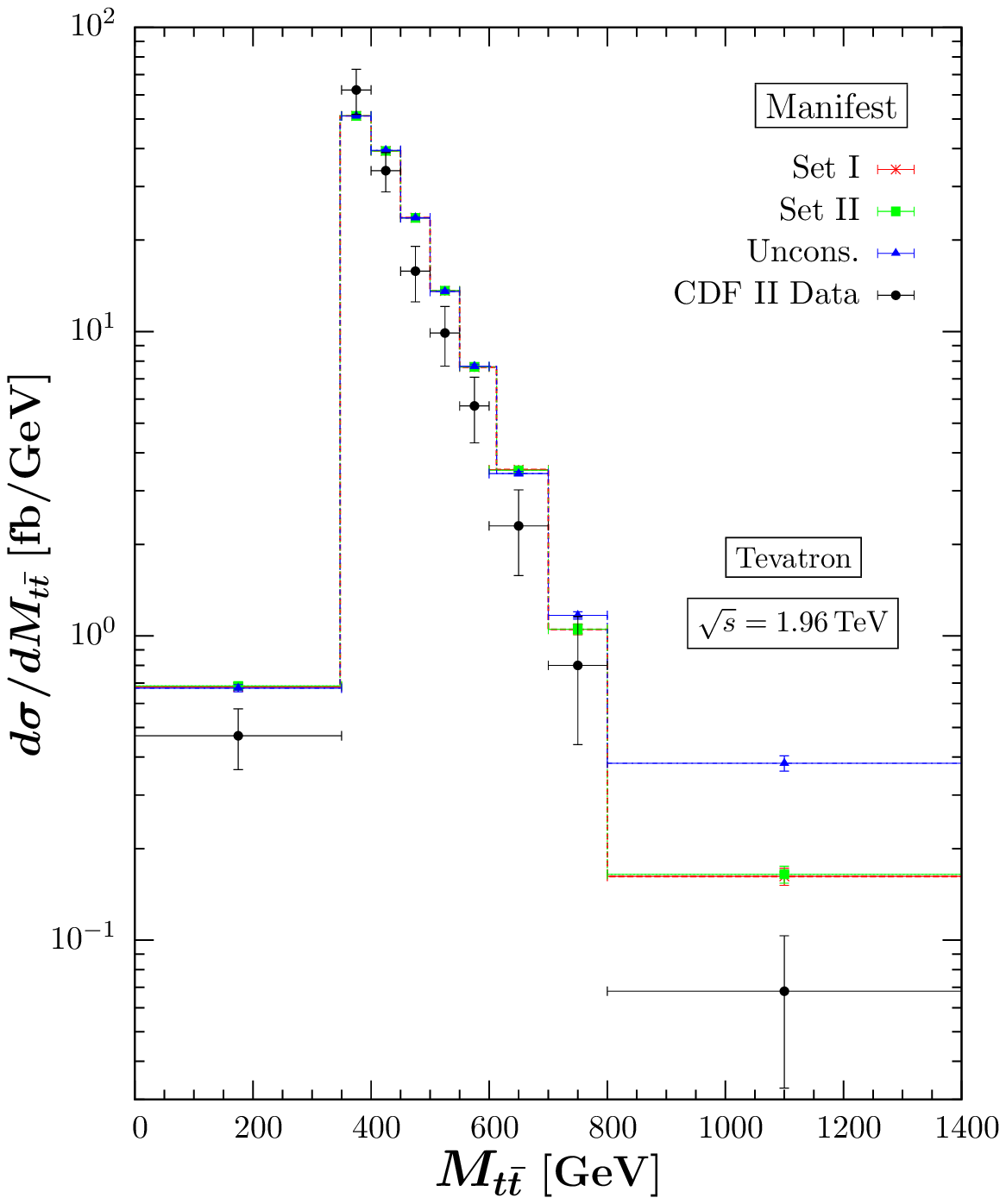} 
$\hspace*{-1cm}
\begin{array}{cc} & \\
\includegraphics[width=3.0in,height=3.0in]{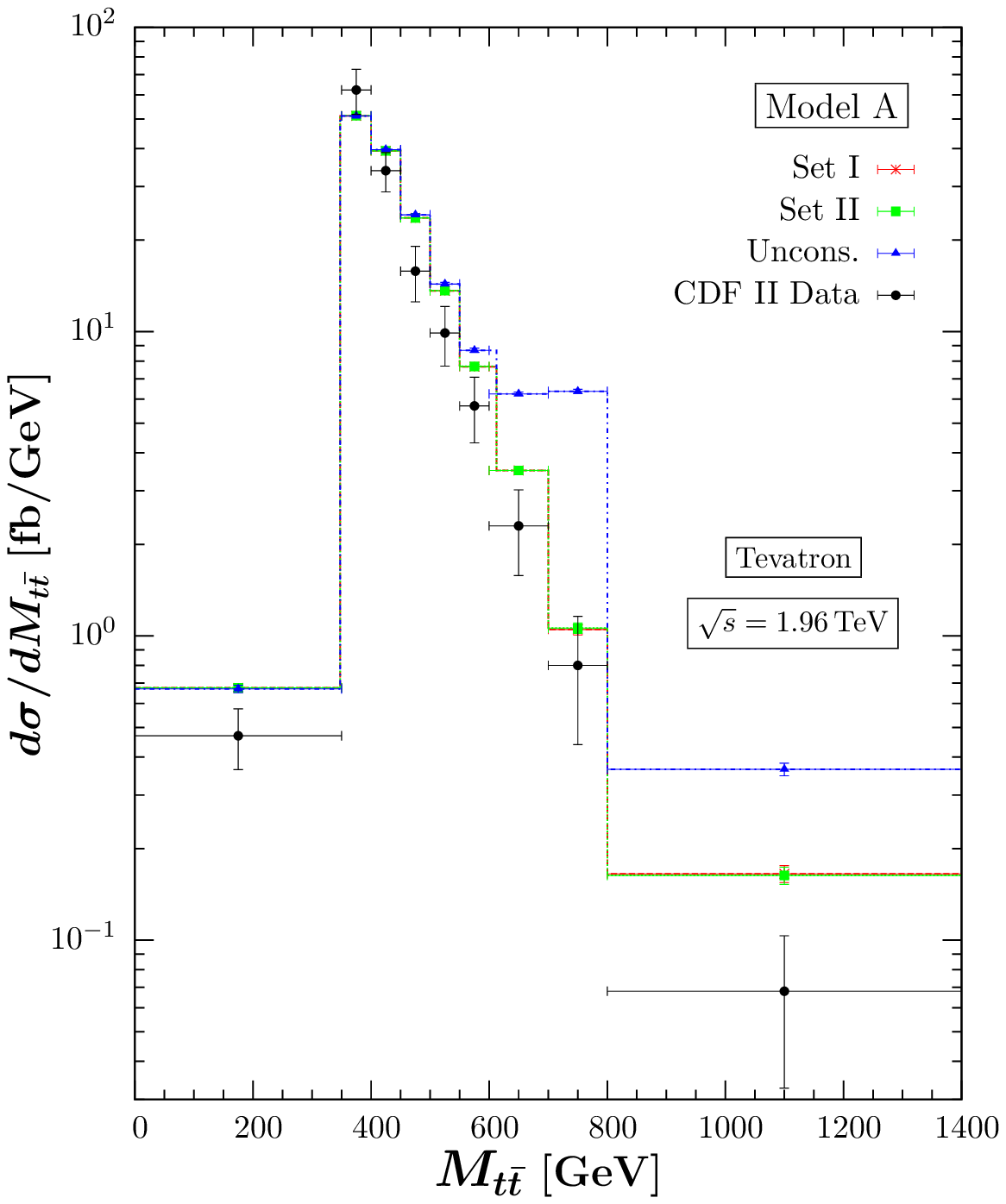} &
\includegraphics[width=3.0in,height=3.0in]{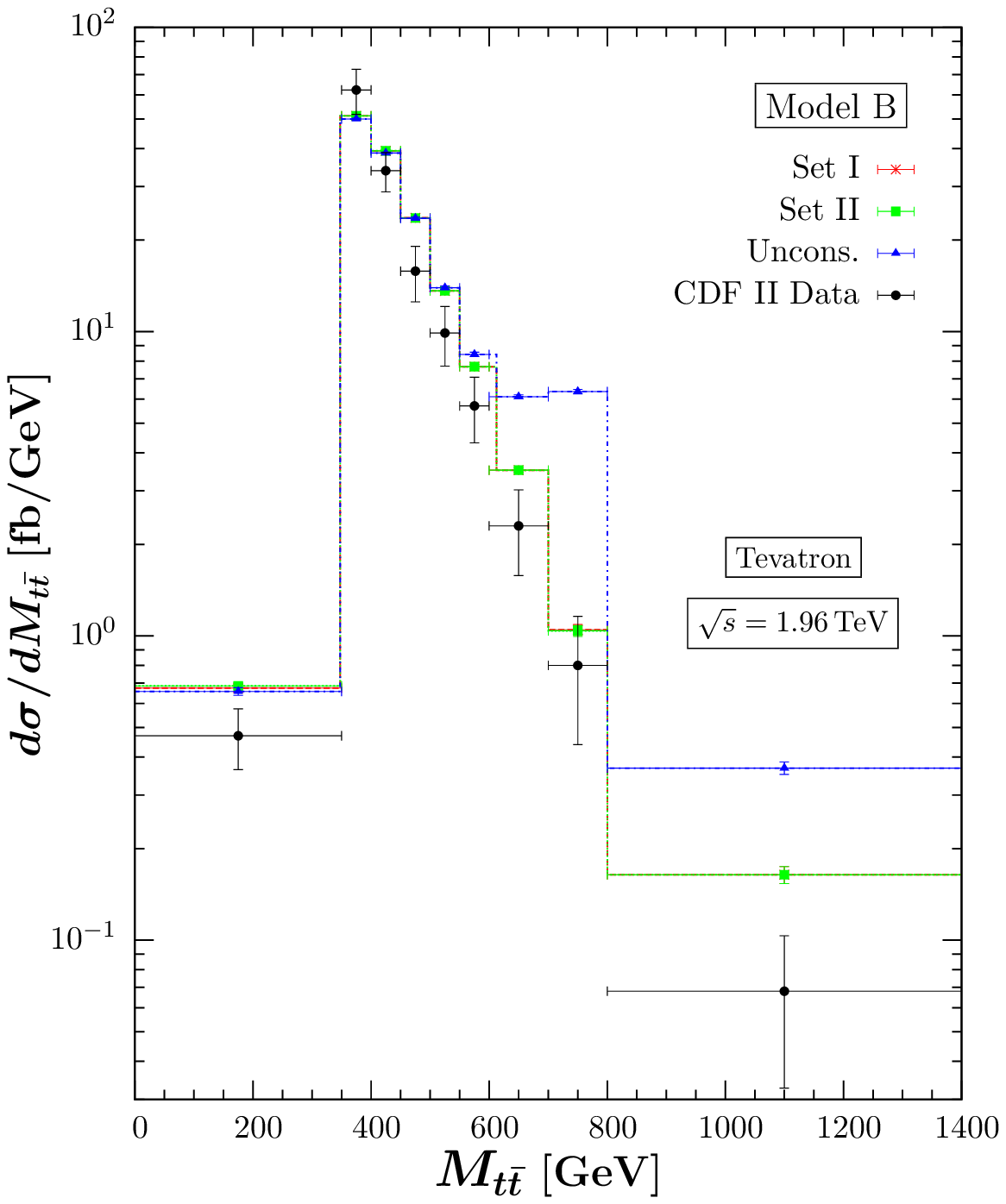}
\end{array}
$  
\end{center}
\vskip -0.5cm
\caption{$t\bar{t}$ invariant mass distribution of differential cross section in Manifest LR model (upper panel), Model A
(lower left panel) and Model B (lower right panel) in comparison with CDF II 5.3 fb$^{-1}$ data. Parameter sets (Set I, Set
II and Unconstrained Set) for each model are given in the Table \ref{params}.  We include as well the uncertainties in
our numerical evaluation. } 
\label{fig:prodxsec-tev}
\end{figure}

The CDF and D0 results impose that in addition to the total production cross section of $t\bar{t}$, the differential cross
section with respect to the invariant mass of  $t\bar{t}$ should also agree with the SM prediction. Thus, in
Fig.~\ref{fig:prodxsec-tev} we graph the  differential cross sections in LR models with respect to the $t\bar{t}$ invariant
mass distributions and compare our calculation with the  CDF II measurement. In the three panels of
Fig.~\ref{fig:prodxsec-tev} we show in sequence the differential cross sections for the Manifest, Model A and Model B for the
two parameter sets Set I (red), Set II (green) as well as the Unconstrained Set (blue).  We include in the figure the 
uncertainties in our calculations, as given by the numerical routine.  The CDF data is given as black lines, and  includes
uncertainties in each bin. Note that care must be taken when comparing the new physics cross-sections against the SM
cross-section, as the selection efficiencies for NP models can be lower. The predicted NNLO SM cross-section requires a SM
$K$-factor of 1.3, while the NNLO corrections to the new physics have not been calculated, so any comparison between the
observed cross-section and the $t {\bar t}$ production cross-section is subject to some uncertainty \cite{Gresham:2011fx}.
Comparing our results to the central value of the combined CDF  $t {\bar t}$ production cross-section to the cross-section of
SM plus new physics for all three  parameters sets show fairly good agreement with the $M_{t{\bar t}}$ distribution measured
by CDF II,  and given our comments above, it probably may yield even better agreement. Thus we insured that,  for the
parameters chosen, both the total and the differential cross sections are consistent with the data. Note however the slight
enhancement of the differential cross section in the Unconstrained set for $M_{t {\bar t}} >500$ GeV, due to low  $M_{Z_R}
=734$ GeV  for Models A and B. The raise is shifted and (not seen due to an uneven bin choice) for the Unconstrained set of
the Manifest model, where $M_{Z_R} =837$ GeV.

We proceed next by examining the asymmetry in the production and decays of the $t\bar{t}$ system. The forward-backward
asymmetry of top quark pairs (${\cal A}_{FB}^{t\bar{t}}$) in $p\bar{p}$  collisions is seen as a precision test of the SM.
The $t\bar{t}$ pair production in SM at the lowest order is symmetric under charge conjugation. At NLO, the interference of
QCD processes involving initial and final state gluon emission $q {\bar q} \to t {\bar t}g$ and $qg \to t {\bar t}q$ will
exhibit a small forward-backward asymmetry. The NLO calculations in the SM yield an asymmetry due to virtual corrections
arising from interference effects, which are opposite in sign and larger than the real emission component.

The forward-backward asymmetry is defined in terms of top quark rapidities as
\begin{eqnarray}
{\cal{A}}_{FB} = \frac{N(\Delta_y>0) - N(\Delta_y<0)}{N(\Delta_y>0) + N(\Delta_y<0)} 
\end{eqnarray}
where $\Delta_y=y_t - y_{\bar{t}}$ is the difference of top and anti-top rapidities and $N$ is the number of events in the 
forward ($\Delta_y>0$) and backward ($\Delta_y<0$) regions. While the cross sections measured by CDF and D0 agree with the SM
expectations, the measured asymmetries deviate   from the NLO SM calculation, by as much as 50\% in the large $M_{t {\bar
t}}$ invariant mass bin. It is the challenge of any new BSM to generate the asymmetry without disturbing the cross section;
it is our intention to verify if this is possible for a realistic left-right model.

We proceed as follows. Since the kinematical cuts in Tevatron analysis are very restrictive, we generate 5 million signal
events in order to minimize the statistical errors. We generate events with {\tt CalcHEP 3.1} \cite{Pukhov:2004ca} using {\tt
CTEQ6M} PDFs. The factorization and renormalization scales $\mu_F = \mu_R = m_t$ are used, and we take the top quark mass
$m_t=172.5$ GeV. We use {\tt Pythia 6.4.18} \cite{Sjostrand:2006za} for showering and {\tt PGS 4} \cite{conway} for jet
reconstruction, $b$-tagging and a rough detector simulation. 
\vspace{0.4cm}
\begin{figure}[htb]
\begin{center}
\includegraphics[width=3.2in,height=2.2in]{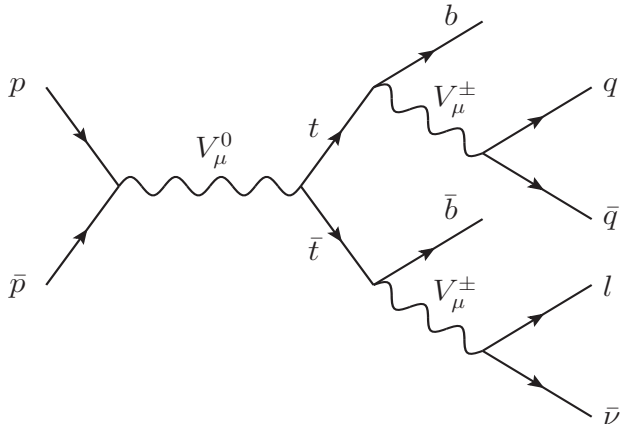}  
\end{center}
\vskip -0.5cm
\caption{$t\bar{t}$ production and decay topology in hadronic and semileptonic events. $V_\mu^0$ represents neutral gauge
bosons $\gamma,g,Z,Z_R$ and $V_\mu^\pm$ the charged ones, $W_{L}^\pm,W_{R}^\pm$. The diagram with the top quark decaying
hadronically is shown but both possibilities are included. }
\label{fig:ttdecay}
\end{figure}

We start the analysis by producing the $t\bar t$ pair, then decaying top quarks semileptonically and hadronically. We
concentrate our analysis on the  $lepton+jets$ topology, where one top quark decays semileptonically ($t\rightarrow bl\nu$)
and the other hadronically ($t\rightarrow bq\bar{q}^\prime$), as in Fig.~\ref{fig:ttdecay}. We select events with one single
lepton (electron or muon) plus missing energy to account for the associated neutrino and a minimum of 4 jets with one jet
$b$-tagged and with the following kinematical cuts,
\begin{eqnarray}
 |\eta^l| < 1  ~~~&,&~~~ |\eta^j| < 2\ , \nonumber \\
 p_T^l > 20~ {\rm GeV} ~~~&,&~~~ p_T^j > 20~ {\rm GeV}\ , \nonumber \\
 \slashed{E}_T \geq 20 ~{\rm GeV} ~~~&,&~~~ |\eta^b| < 1,
\end{eqnarray}
where $l,j,b$ denote lepton ($e,\mu$), jet ($u,d,c,s$) and $b$-quark parameters, respectively. The jets are reconstructed
using a cone algorithm with $\Delta R=\sqrt{\Delta\phi^2 + \Delta\eta^2}<0.4$. Here $b-$jets, tagged with the loose {\tt
SECVTX} algorithm, are restricted to $|\eta^b|<1$. We used the default $b$-tagging efficiency and functions for Tevatron
given in {\tt PGS 4}. The efficiency of the signal to pass through the cuts (after showering, clustering and detector
simulations) allows only $2\%$ signal events to survive the kinematical cuts to yield  the forward-backward asymmetries.
 
The number of events are scaled to the NNLO cross sections using the standard $K$-factor for the Tevatron. We have calculated
the left-right contribution at the LO (including the LO SM, LR and the interference between the two)  to the asymmetries.
We listed asymmetries obtained for the four different regions, for all models studied, in Table \ref{tbl:asym-tev-nlo}. The
first two rows are parton level asymmetries, the first row obtained by unfolding the CDF data and the second for the MCFM.
The remaining rows compare the CDF signal data to our various models\footnote{In fact, the signal level data for the regions 
$|\Delta_y|\geq 1$ or $|\Delta_y|\leq 1$ are not presented in \cite{newcdf}. So, we have used the data-level values including
the background.}. As it is seen from the Table \ref{tbl:asym-tev-nlo}, the LO left-right contributions to the asymmetries are
relatively small. The results might have been enhanced if the left-right contributions were calculated at the NLO which is
beyond the scope of this work. We have chosen to compare our results, simulated to the final states, with the CDF signal. The
reason is that the errors in the signal results are much smaller than the ones evolved to parton level, and thus this
comparison gives a better measure of the deviation of our results from the data.  We include a reduced $\chi^2$ analysis as a
measure of how well the models perform. As expected all the scenarios other than Unconstrained LR Model A and Model B give
asymmetries more than 4 sigma away from the observed ones. The situation for the Model A and B of the Unconstrained LR is
close to 1 sigma.

\begin{table}[b]
\begin{tabular*}{0.99\textwidth}{@{\extracolsep{\fill}} ccccccc} 
\hline\hline
& & $A_{FB}^{t\bar{t}}$ & $A_{FB}^{t\bar{t}}$ & $A_{FB}^{t\bar{t}}$ & $A_{FB}^{t\bar{t}}$ & $\chi^2_{red}$\\ 
& & $|\Delta_y|<1$ & $|\Delta_y|\geq 1$ & $M_{t\bar{t}}<450$ GeV & $M_{t\bar{t}}\geq 450$ GeV & $(4\, \rm d.o.f.)$\\ \hline
\multicolumn{2}{c}{CDF(parton-level)}  & $~0.026\pm 0.118~$ & $~0.611\pm 0.256~$ & $-0.116\pm 0.153~$ & $~0.475\pm 0.114~$ &
\\
\multicolumn{2}{c}{MCFM(parton-level)}   & $0.039\pm 0.006$ & $0.123\pm 0.008$ & $0.040\pm 0.006$ & $0.088\pm 0.013$ & \\
\multicolumn{2}{c}{CDF(signal-level)}  & $~0.021\pm 0.031~$ & $~0.208\pm 0.062~$ & $-0.022\pm 0.043~$ & $~0.266\pm 0.062~$ &
\\ \hline
\multicolumn{1}{c}{\multirow{6}{*}{~LR~}} & Manifest-I & $0.0025$ & $0.0174$ & $0.0030$ & $0.0086$ & $6.8$\\ 
 & Manifest-II & $0.0098$ & $0.0162$ & $0.0091$ & $0.0137$ & $6.7$\\ 
\multicolumn{1}{c}{}
 & Model A-I & $0.0063$ & $0.0143$ & $0.0065$ & $0.0096$ & $6.9$\\ 
\multicolumn{1}{c}{}
 & Model A-II & $0.0043$ & $0.0131$ & $0.0051$ & $0.0072$ & $7.0$\\ 
\multicolumn{1}{c}{}
 & Model B-I & $0.0077$ & $0.0121$ & $0.0062$ & $0.0118$ & $6.9$\\ 
\multicolumn{1}{c}{}
 & Model B-II & $0.0035$ & $0.0038$ & $0.0029$ & $0.0044$ & $7.3$\\ \hline
\multicolumn{1}{c}{\multirow{3}{*}{$\begin{array}{c} \text{Uncons.} \\ \text{LR} \end{array}$}} & Manifest & $0.0065$ &
$0.0280$ & $0.0024$ & $0.0222$ & $6.1$\\ 
\multicolumn{1}{c}{}
 & Model A & $0.0532$ & $0.2400$ & $0.0078$ & $0.1832$ & $0.9$\\ 
\multicolumn{1}{c}{}
 & Model B & $0.0444$ & $0.2189$ & $-0.0084$ & $0.1751$ & $0.7$\\ \cline{1-7} \hline
 \hline
\end{tabular*} 
\caption{The Forward-Backward Asymmetry at the Tevatron in the  LR models: Manifest, Model A and Model B, compared
with the CDF data. We include, in the first two rows, the unfolded CDF results and the MCFM calculation. Parameter sets 
(SetI, Set II and Unconstrained) for each model are given in the Table \ref{params}.}
\label{tbl:asym-tev-nlo}
\end{table}

It is apparent from Table \ref{tbl:asym-tev-nlo} that, while the models yield a slightly enhanced or slightly suppressed 
forward-backward asymmetry with respect to the SM in one region or another, none of the phenomenologically viable LR models
can reproduce large enough anomaly seen at the Tevatron. The results for the benchmark points chosen for Manifest, Model
A and Model B are however fairly consistent in the both sign and size of the anomaly. Moreover the asymmetry seems to depend
sensitively on $M_{W_R}$ and on the ratio $g_R/g_L$, but to a lesser extent on $\sin \alpha$, the measure of flavor violation
in the right-quark sector. From our previous investigations of the parameter space we know that $M_{W_R}$ and $g_R/g_L$ are
closely correlated, as a decrease or increase  in one forces a decrease or increase in the other to satisfy low energy
constraints. We are thus confident that results for the sets chosen are a true indication of LR model predictions. As sets I
and II represent very different regions of the parameter space, and different variants of the model, this is further
confirmation that our results are robust and do not depend on the specific points chosen in the parameter space of the LR
model. One can obtain higher asymmetries  (consistent with the data) in a LR model not subjected to experimental
constraints (last two rows in Table \ref{tbl:asym-tev-nlo}), as indeed is the case for models constructed specifically to
explain the asymmetry.  

\begin{figure}[htb] 
\begin{center}
$\hspace*{-0.5cm}
\begin{array}{ccc}
\includegraphics[width=2.15in,height=2.4in]{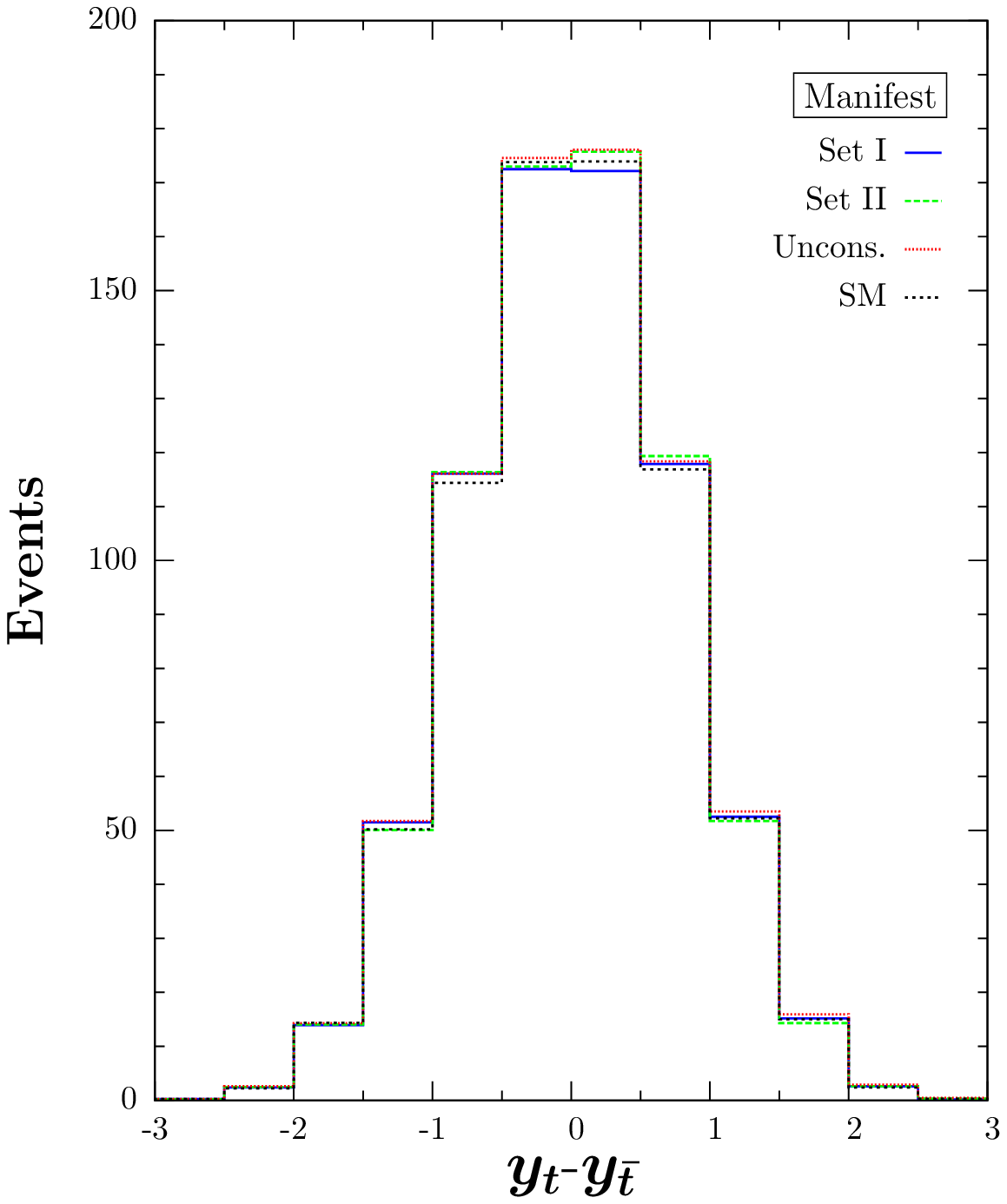} &
\includegraphics[width=2.15in,height=2.4in]{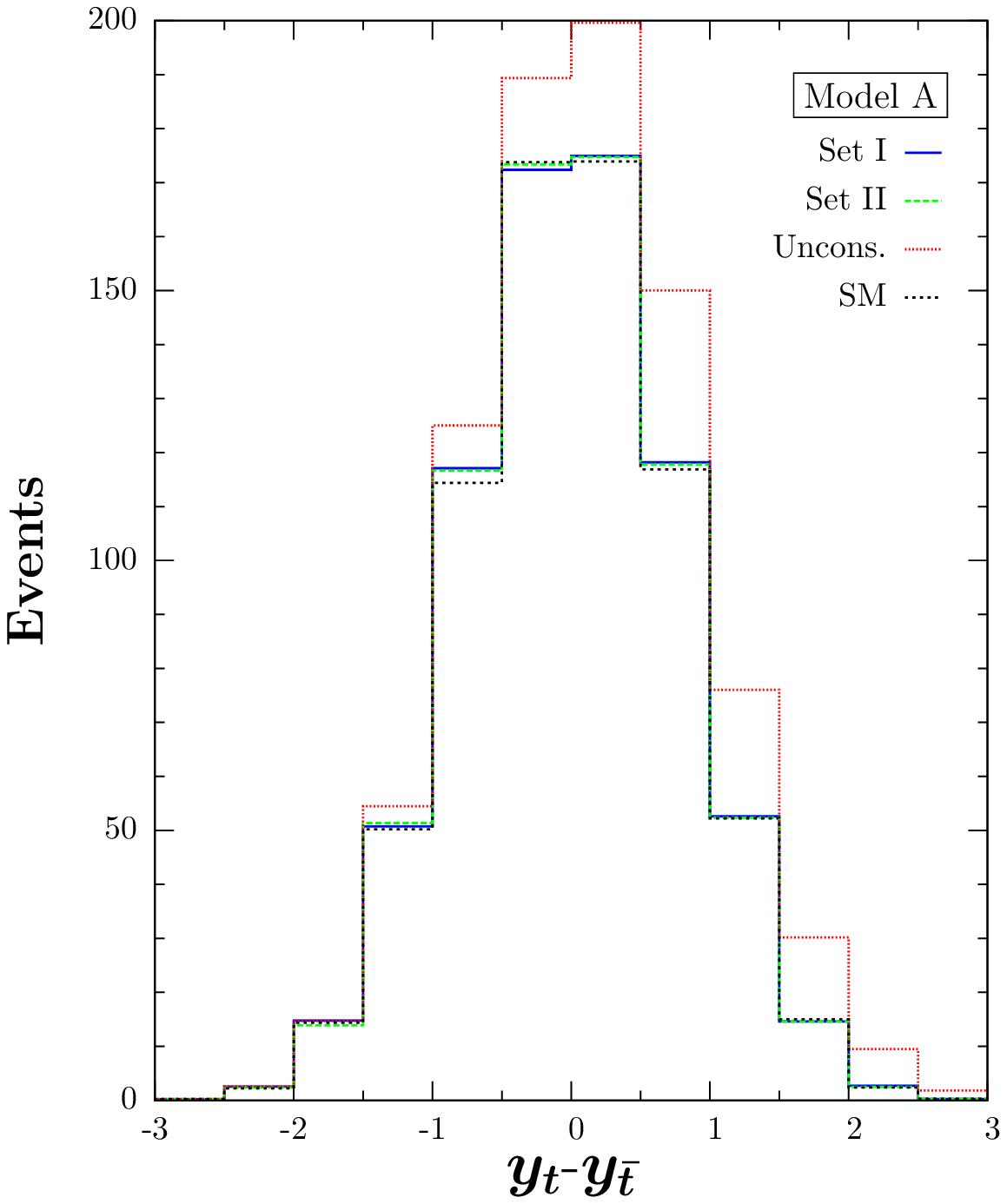} &
\includegraphics[width=2.15in,height=2.4in]{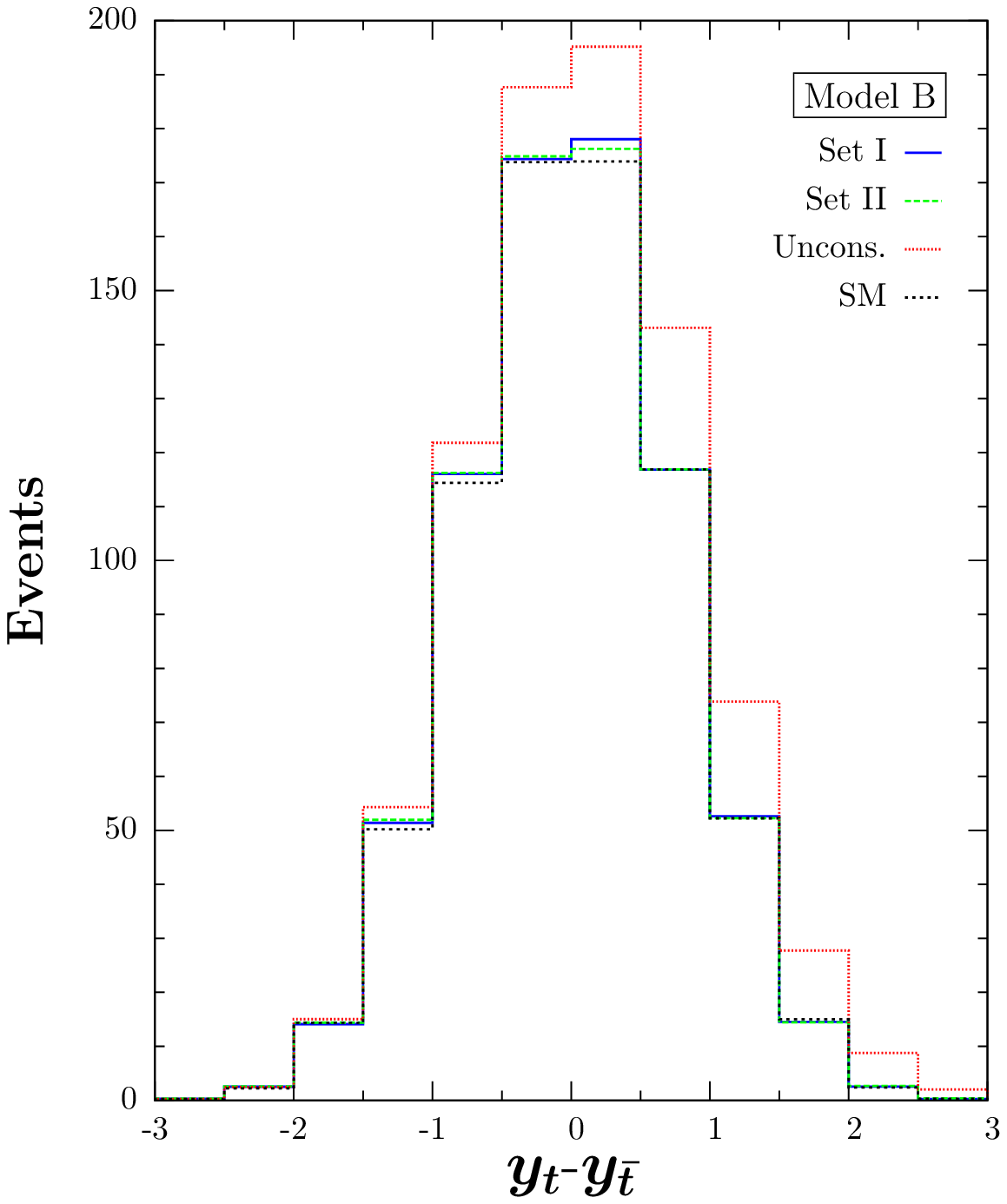} \\
\includegraphics[width=2.15in,height=2.4in]{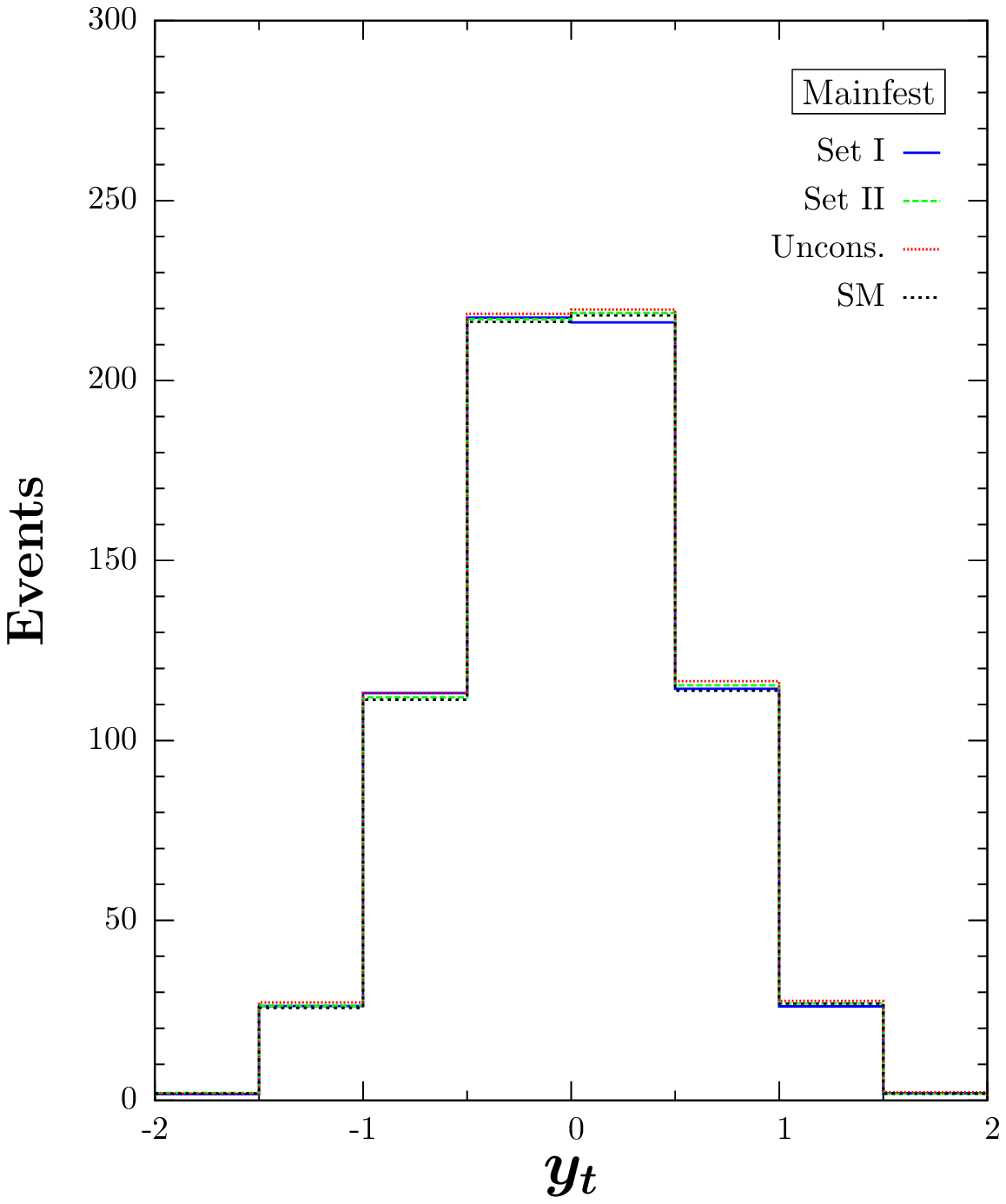} &
\includegraphics[width=2.15in,height=2.4in]{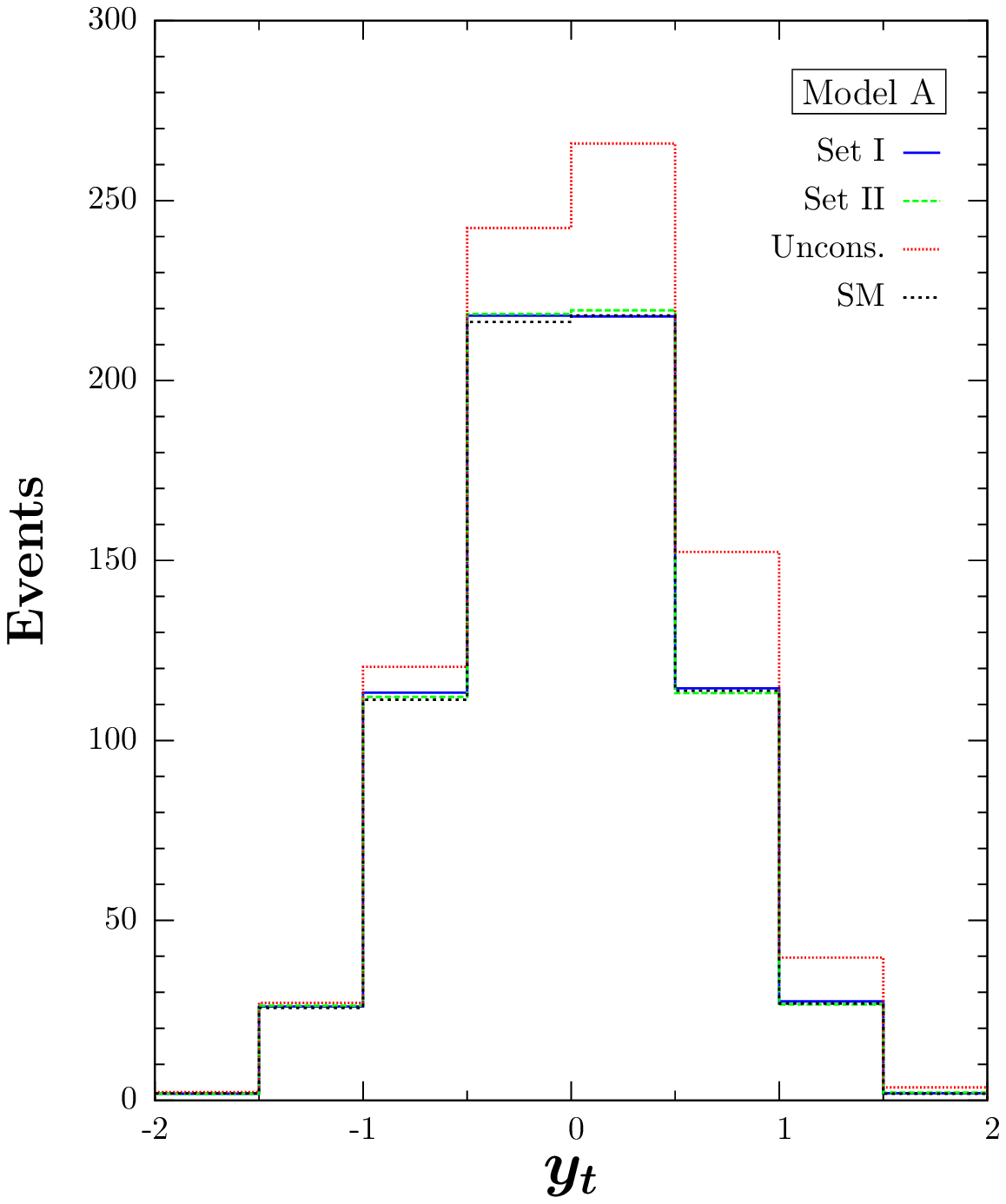} &
\includegraphics[width=2.15in,height=2.4in]{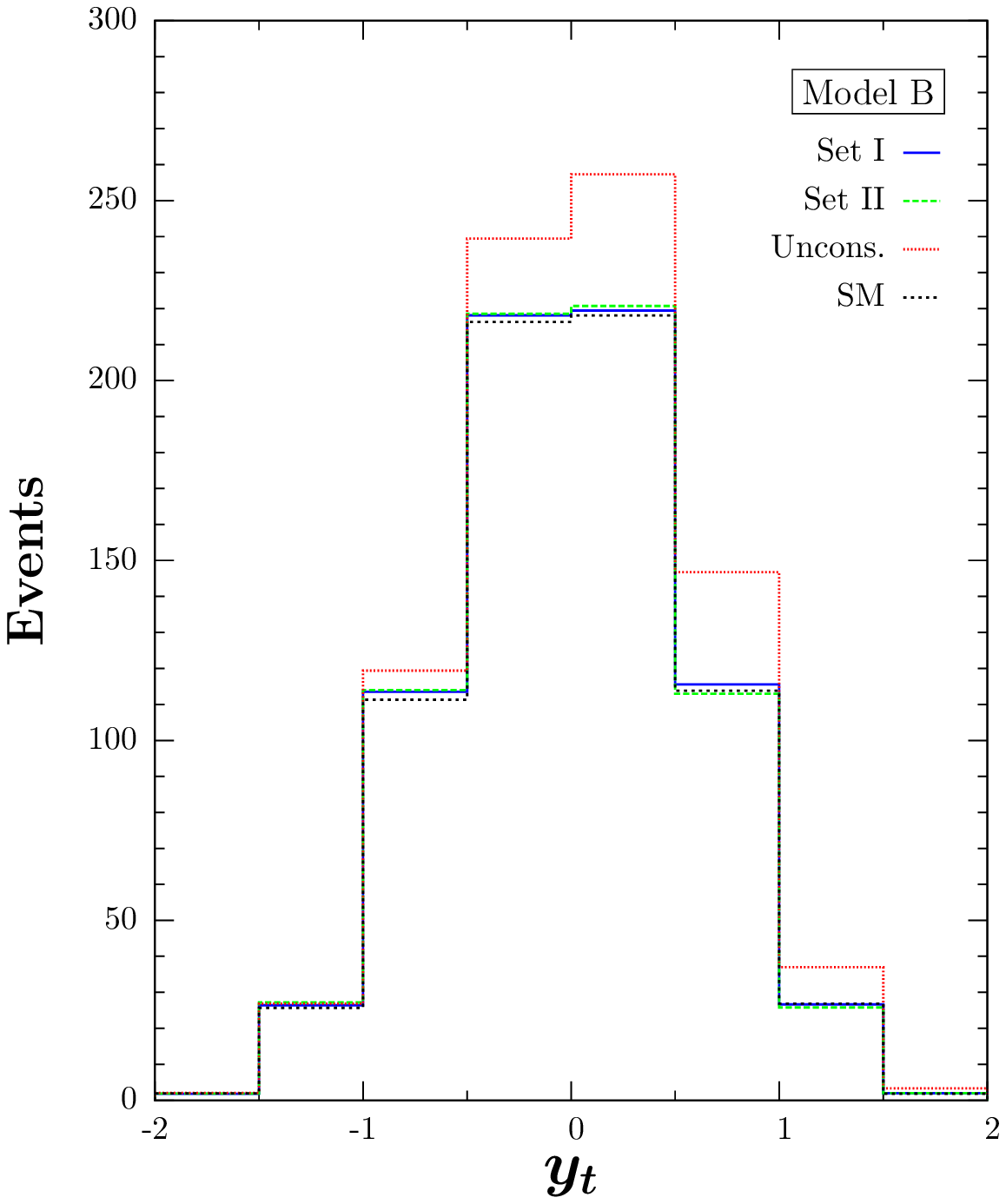}
\end{array}
$ 
\end{center}
\vskip -0.5cm
\caption{$\Delta_y$(upper row) and $y_t$(lower row) distributions in Manifest LR model (left panel), Model A (middle panel)
and Model B (right panel) at the Tevatron. 
Parameter sets (Set I, Set II and Unconstrained) for each model are given in the Table \ref{params}.} 
\label{fig:rapid-tev}
\end{figure}
We proceed to investigate the features of the signal in LR models. In Fig.~\ref{fig:rapid-tev} we show the distributions of
rapidity differences $\Delta_y$ in the upper row, and top quark rapidity $y_t$ in lower row, for three different LR models.
In order to generate a  large asymmetry in the high invariant mass bin, the rapidity must be increased and skewed
significantly with respect to the SM distribution. Additional high mass gauge bosons could sometimes produce this effect. We
show the Manifest model (left panel), Model A (middle panel) and Model B (right panel) with Set I (blue) Set II  (green) the
Unconstrained (red) and the SM (black) in each panel. We did not perform a global fit to the data, as our results do not
agree with the CDF measurements.  The
results are however consistent among the different models obeying low energy constraints, and parameters sets chosen, at
least making the left-right model very predictable.

\begin{figure}[htb] 
\begin{center}
$\hspace*{-0.6cm}
\begin{array}{ccc}
\includegraphics[width=2.15in,height=2.4in]{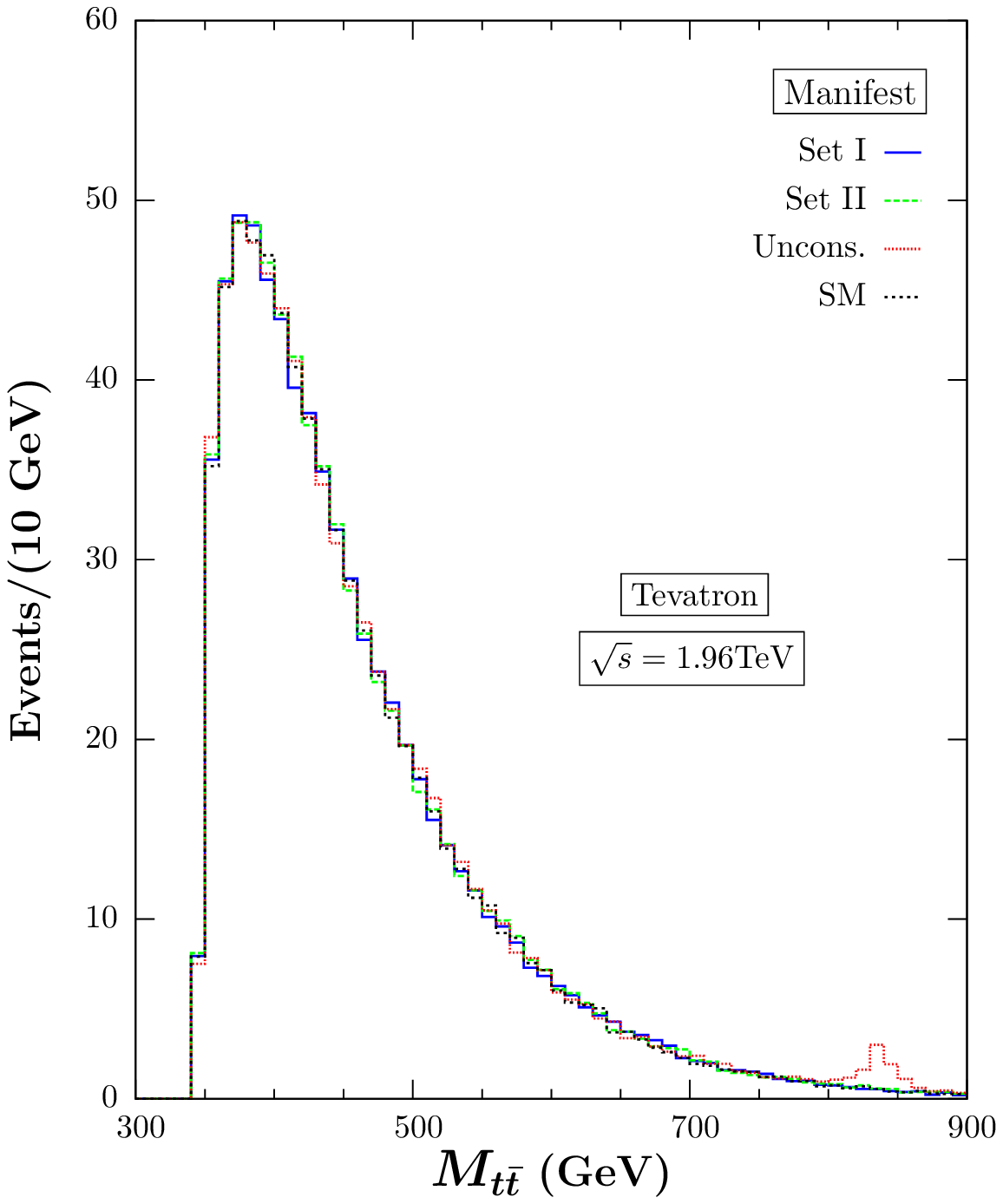} &
\includegraphics[width=2.15in,height=2.4in]{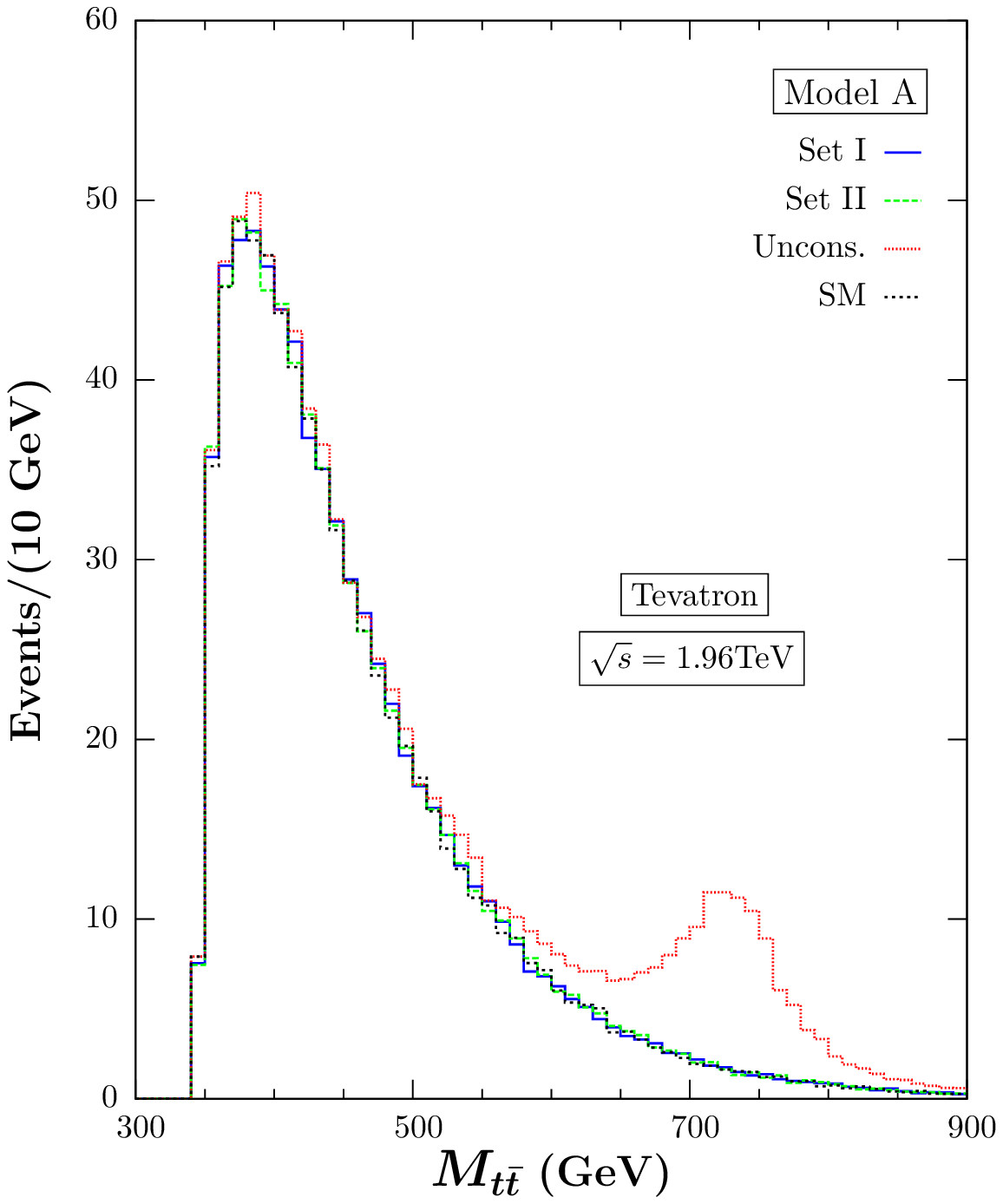} &
\includegraphics[width=2.15in,height=2.4in]{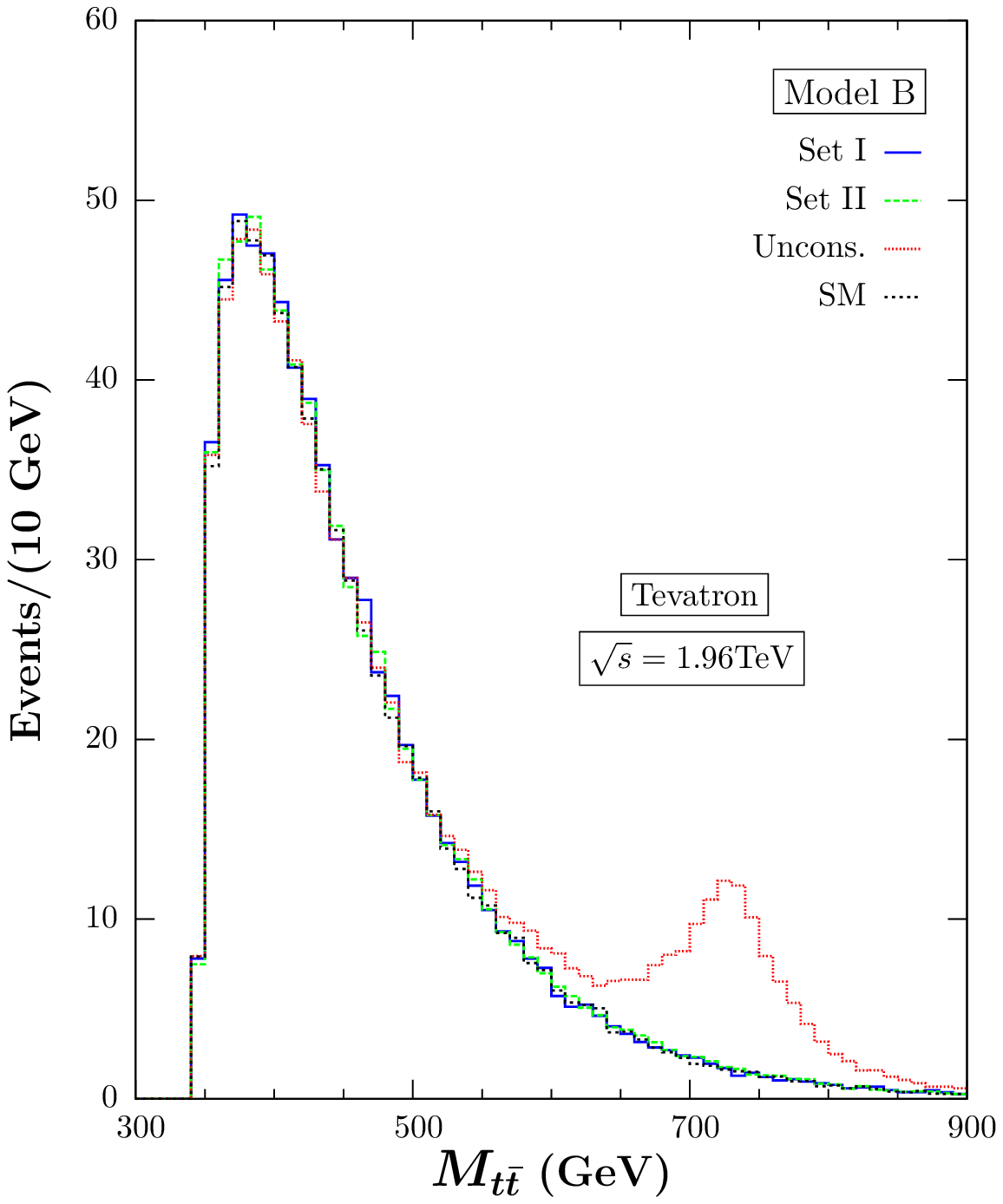}
\end{array}
$ 
\end{center}
\vskip -0.5cm
\caption{$t\bar{t}$ invariant mass distributions at the Tevatron in Manifest LR model (left panel), Model A (middle panel)
and Model B (right panel) in comparison with the SM. Parameter sets (Set I, Set II and Unconstrained) of each model are given
in the Table \ref{params}.} 
\label{fig:minv-tev}
\end{figure}

In Fig.~\ref{fig:minv-tev} we give invariant mass distributions in {\tt Pythia} of LR models at Tevatron, for the Manifest LR
model (left panel), Model A (middle panel) and Model B (right panel). The number of events are scaled to NNLO cross-sections
with standard $K$-factor. Comparison with the SM expectations again shows consistency.

Both this figure and the previous one show that realistic LR models, which obey low-energy constraints, cannot yield the
measured CDF asymmetry. The Unconstrained model shows an increase in the differential cross section, corresponding to a $Z_R$
peak around $734$ GeV in Models A and B, and  a less pronounced one at $837$ GeV in the Manifest left-right case. These are
close to the experimental limit at the Tevatron, and the first two are likely already ruled out. Changing the ratio $g_R/g_L$
and lowering the $W_R$ mass may be able to achieve consistency of left-right models with the asymmetry data, but these models
do not satisfy other phenomenological constraints and are thus unrealistic.

\section{Cross section for $t {\bar t}$ pair production and  asymmetries at the LHC}
\label{sec:LHC}

As the Tevatron results show interesting discrepancies with the SM expectation, it is important to evaluate the asymmetries
and cross sections for $t {\bar t}$ production at the LHC. Naturally one might ask if such a pursuit is worthwhile, as
we have shown in the previous section that the model cannot explain the Tevatron asymmetries. The large forward-backward
asymmetry at the Tevatron, although an exciting signal for new physics, may not arise from new interactions or new particles.
It could arise from a kinematical enhancement of the $t {\bar t}$ pair, or from a hidden sector. Even the experimental
situation at the Tevatron is not yet clear, as the errors on the measurements are significant; also the CDF results show a
strong mass dependence of the asymmetry not confirmed by the  D0 measurements. At LHC different production mechanisms
dominate and other asymmetries are at play. Measurements of the charge asymmetry at CMS and ATLAS at the LHC (which appear to
be small and negative, though the uncertainties are too large to make a firm statement) are hard to reconcile with the
Tevatron results. Predictions for both colliders are important to understand the dynamics of different gauge symmetries and
their effect on different asymmetries. This is particularly interesting for our model, which can reproduce the Tevatron
cross section but not the asymmetry. The natural question is: what is the prediction for the LHC?  While the Tevatron has
collected about a thousand tops, the LHC, even with ${\cal L}=1$ fb$^{-1}$ has amassed almost an order of magnitude more,
making the errors in the production cross section at $\sqrt{s}=7$ TeV already competitive with those at the Tevatron with
${\cal L}=5.3$ pb$^{-1}$, while the invariant mass $M_{t {\bar t}}$ investigated extends to 2.5 TeV (with 200 pb$^{-1}$),
versus 1.8 TeV for the Tevatron. LHC will provide measurements of top quark properties, shedding light on models on NP and
electroweak symmetry breaking. Agreement or disagreement with this data would open (or perhaps narrow) questions to do with
the validity or restrictions of the model. For example, the CMS Collaboration has recently presented the first measurement of
charge asymmetry in $t {\bar t}$ production \cite{CMSasym}
\begin{eqnarray}
A^\eta_C&=-&0.016 \pm 0.030 ({\rm stat}) ^{+0.010}_{-0.019}  ({\rm syst})\nonumber \\
A^y_C&=-&0.013 \pm 0.026 ({\rm stat}) ^{+ 0.026}_{-0.021} ({\rm syst})
\end{eqnarray}
The first one based on pseudorapidities ($\eta$), the second on the rapidity ($y$) of the two top quarks, while the combined
($e+jets$ and $\mu+jets$ channels) ATLAS \cite{ATLASasym} result is 
\begin{eqnarray}
A_C&=&-0.024 \pm 0.016({\rm stat}) \pm 0.023 ({\rm syst)}
\end{eqnarray}
As seen, these results have so far large statistical uncertainties, but this uncertainty is expected to decrease with more
data, while the systematic one will improve with improved detector simulation. 

The Tevatron however is a better machine for measuring the forward-backward asymmetry. At the Tevatron, the forward-backward
asymmetry measures the tendency of the top quark (in the $t {\bar t}$ pair) to move along the direction of the incoming
quark rather than along the direction of the incoming anti-quark. At LHC, the measurement of any asymmetry is very subtile.
Its charge-symmetric initial state ($pp$, or the dominant $gg, qq$ partonic level channels) does not provide a framework to
differentiate between initial partons in the $t {\bar t}$ production. To define an asymmetry one must rely on subleading
contributions to the $t {\bar t}$ production cross section from $q {\bar q}$ and $qg$, with different partons in the initial
state. In this case, the forward backward asymmetry  represents a charge asymmetry in the decay $q{\bar q}, qg \to t {\bar
t}+X$ \cite{Hewett:2011wz}, though several other types of asymmetries have been defined \cite{Zhou:2011dg} and used to 
discriminate between BSM models.

We proceed to analyze the properties of the left-right model in top pair production and decays. We evaluate the $t{\bar t}$
production at the LHC following the same procedure used in the previous section to analyze the signal at the Tevatron. First,
we estimate the total and differential cross section for $t {\bar t}$ production for the models under investigation, then we
proceed to define and analyze the charge asymmetry. 

At the LHC, the $t\bar{t}$ production is dominated by gluon fusion in $pp$ collisions. In our calculation we implement the
models in {\tt CalcHEP 3.1} for the evaluation of  production cross sections at LO level. We normalize the cross sections to
NNLO using the NNLO K-factor ($K=1.6$ for LHC) and we present them in Table \ref{tbl:prodxsec-lhc} for both $\sqrt{s}=7$ TeV
and $\sqrt{s}=14$ TeV, for the same parameter sets and models as discussed in the previous section and given explicitly in
Table \ref{params}. While the SM and Manifest LR model are completely consistent for both Set I and Set II parameters, Models
A and B predict a slightly smaller (about 8\%) production cross section (but consistent for both Set I and II), all of which
agree with the measured value (including errors) at ATLAS at $\sqrt{s}=7$ TeV \cite{Aad:2010ey} and with the SM predictions
at NNLO \cite{Cacciari:2008zb},
\begin{eqnarray}
\sigma_{t {\bar t}}^{ATLAS}&= &145 \pm 31^{+42}_{-27}~{\rm pb},\nonumber\\
\sigma_{t {\bar t}}^{NNLO}&= &150 ~{\rm pb}
\end{eqnarray}
while the prediction  for the cross section in the SM at NNLO at $ \sqrt{s}=14$ TeV is $\sigma_{t {\bar t}}^{NNLO}= 919\pm 4
$ pb \cite{Cacciari:2008zb}. A complete analysis of the production cross section should include subsequent decays of the top
quark, as only a detailed analysis would be able to conclude if one can distinguish various scenarios. We present below some
details of our analysis. 


\begin{table}[htb]
\begin{tabular*}{0.98\textwidth}{@{\extracolsep{\fill}}cccc} 
\hline\hline
{\bf SM} & & & \\ 
$\sigma_{\rm 7 TeV}$(pb) & $167\pm 0.17$ & & \\ 
$\sigma_{\rm 14 TeV}$(pb) & $921\pm 1.20$ & & \\ 
{\bf Manifest}& ~{\bf Set I}~ & ~{\bf Set II}~ & {\bf Uncons.} \\ 
$\sigma_{\rm 7 TeV}$(pb) & $168\pm 0.23$ & $168\pm 0.20$ & $169\pm 0.19$ \\ 
$\sigma_{\rm 14 TeV}$(pb) & $924\pm 1.99$ & $923\pm 2.30$ & $926\pm 1.41$ \\ 
{\bf Model A} & ~{\bf Set I}~ & ~{\bf Set II}~ & {\bf Uncons.} \\ 
$\sigma_{\rm 7 TeV}$(pb) & $168\pm 0.12$ & $168\pm 0.14$ & $179\pm 0.11$ \\ 
$\sigma_{\rm 14 TeV}$(pb) & $922\pm 1.33$ & $921\pm 1.46$ & $967\pm 1.82$ \\
{\bf Model B} & ~{\bf Set I}~ & ~{\bf Set II}~ & {\bf Uncons.} \\ 
$\sigma_{\rm 7 TeV}$(pb) & $168\pm 0.15$ & $168\pm 0.12$ & $178\pm 0.10$ \\ 
$\sigma_{\rm 14 TeV}$(pb) & $919\pm 1.31$ & $921\pm 1.04$ & $962\pm 1.52$ \\  \hline \hline
\end{tabular*}
\caption{$t\bar{t}$ production cross-sections at LHC, for both $\sqrt{s}= 7$ TeV and $\sqrt{s}=14$ TeV.}
\label{tbl:prodxsec-lhc}
\end{table}

\begin{figure}[t] 
\begin{center}
$\hspace*{-1.1cm}
\begin{array}{ccc}
\includegraphics[width=2.35in,height=2.4in]{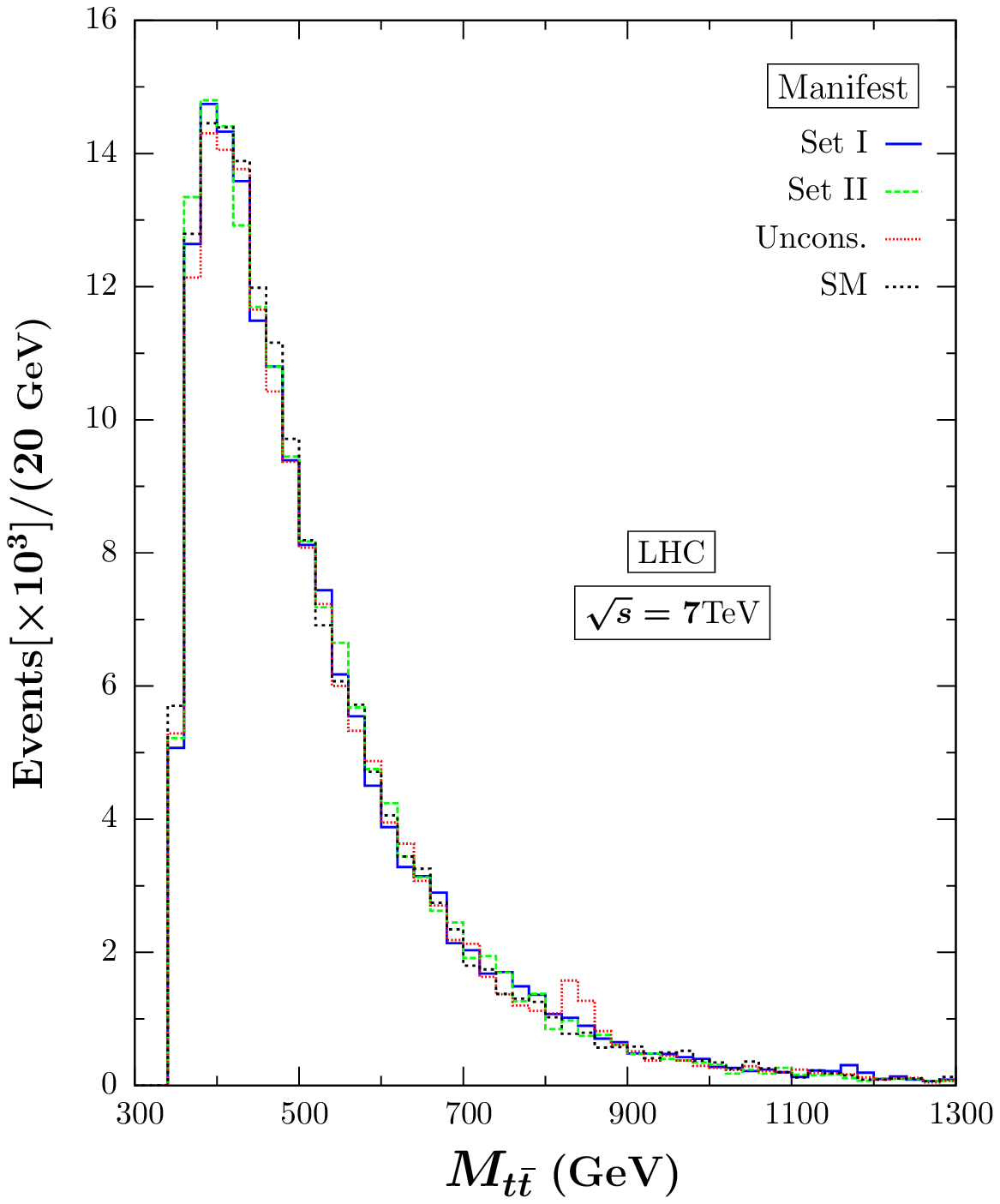} \hspace*{-0.5cm}&
\includegraphics[width=2.35in,height=2.4in]{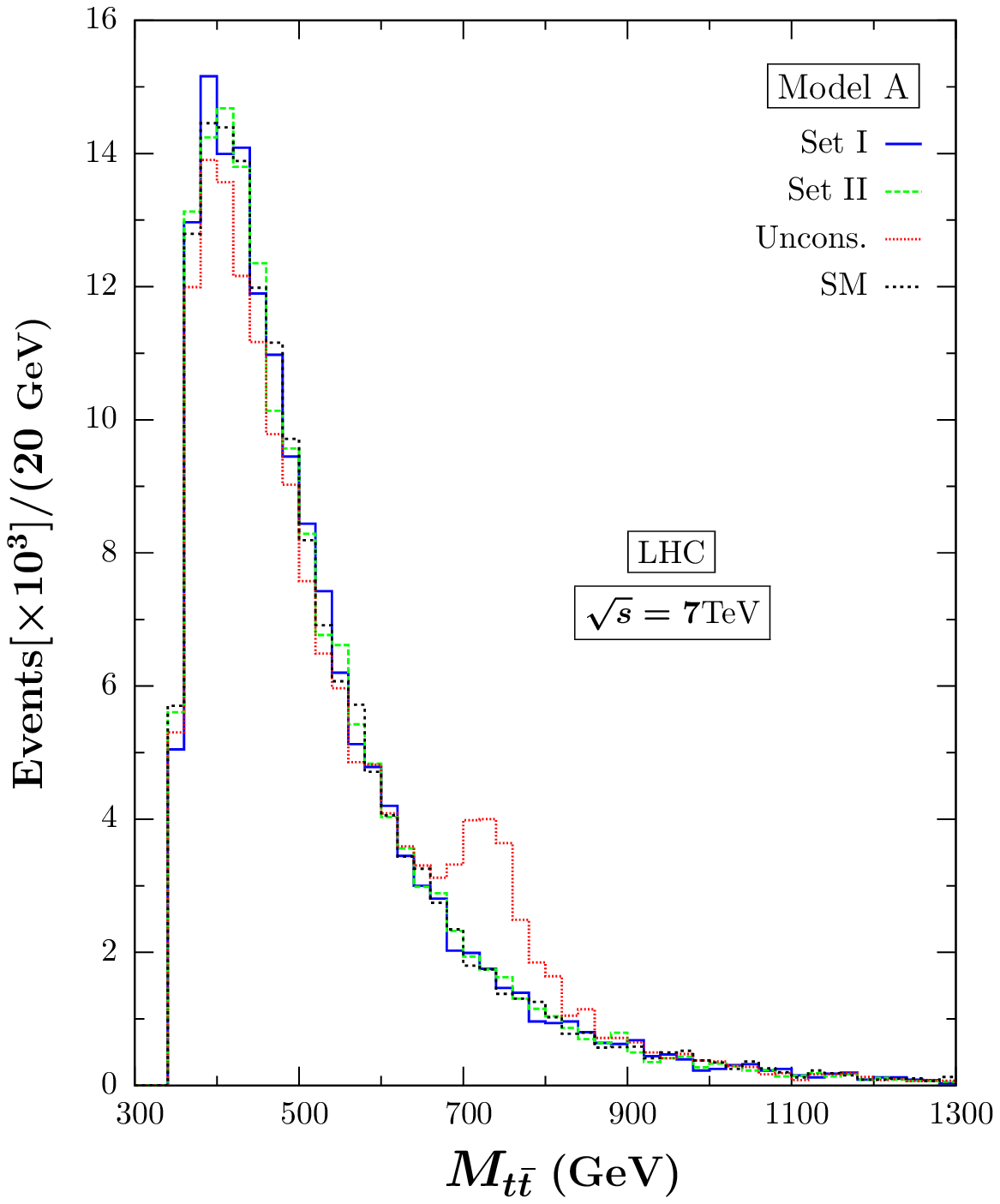}\hspace*{-0.5cm}&
\includegraphics[width=2.35in,height=2.4in]{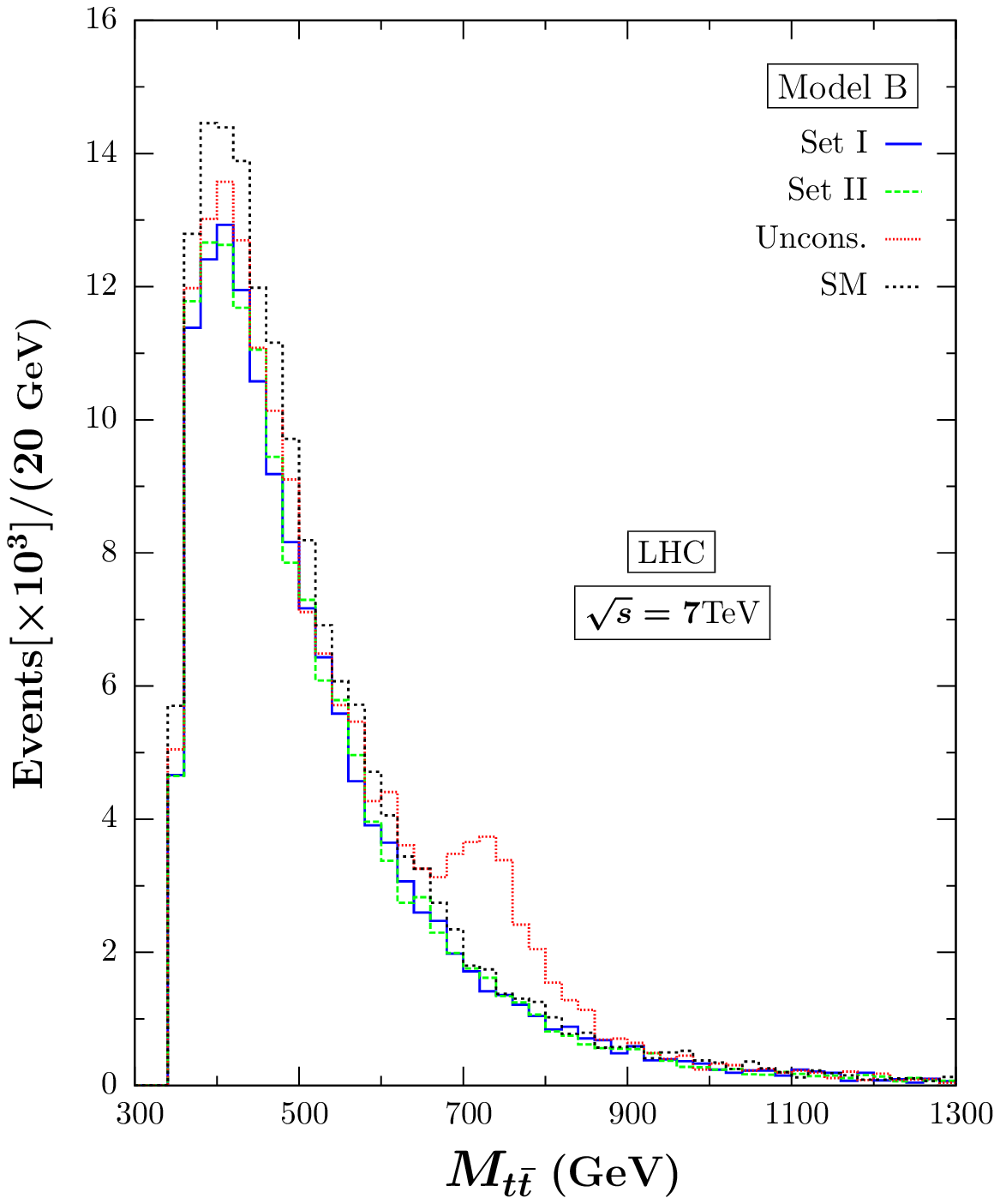}\\
\includegraphics[width=2.35in,height=2.4in]{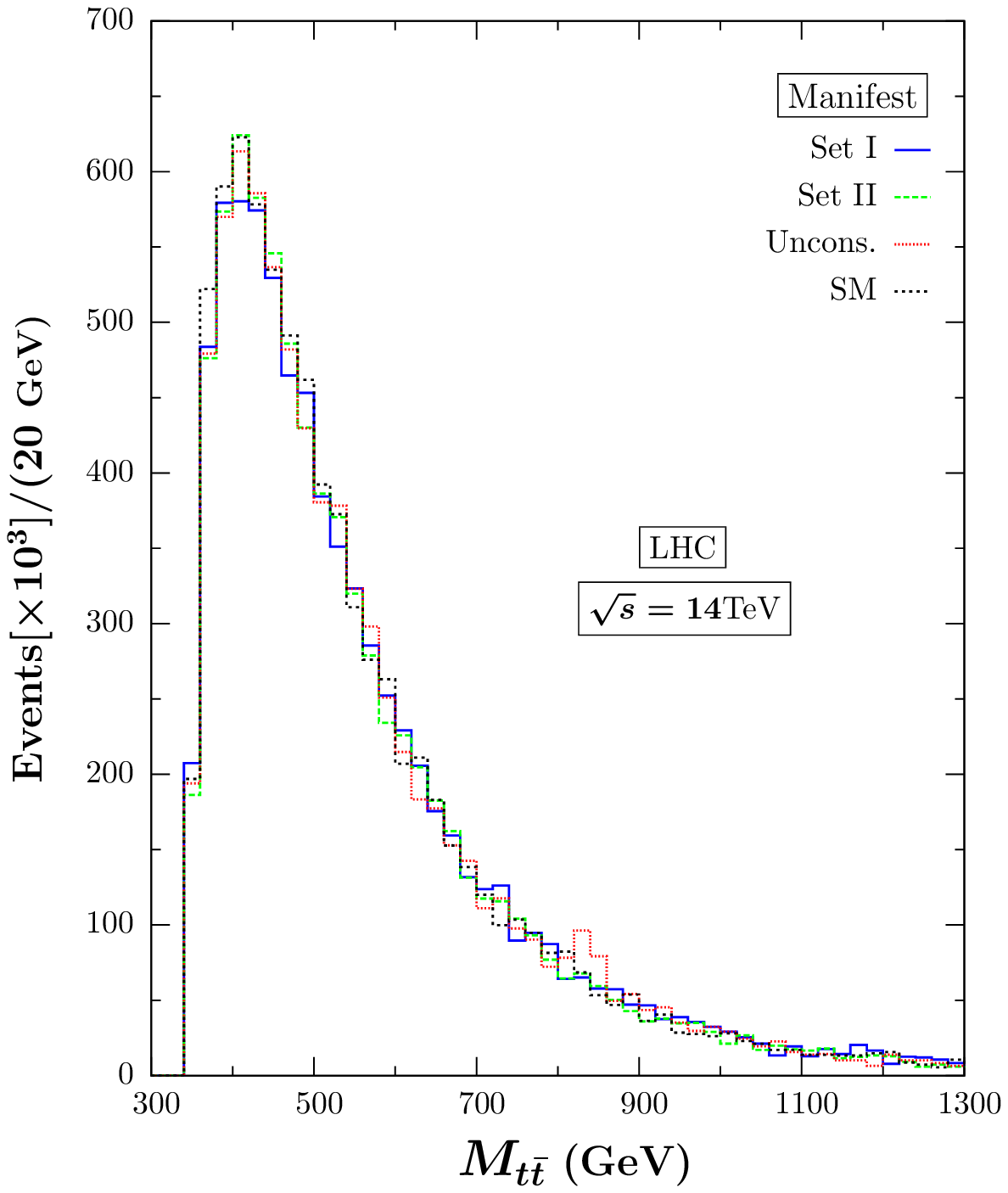} \hspace*{-0.5cm}&
\includegraphics[width=2.35in,height=2.4in]{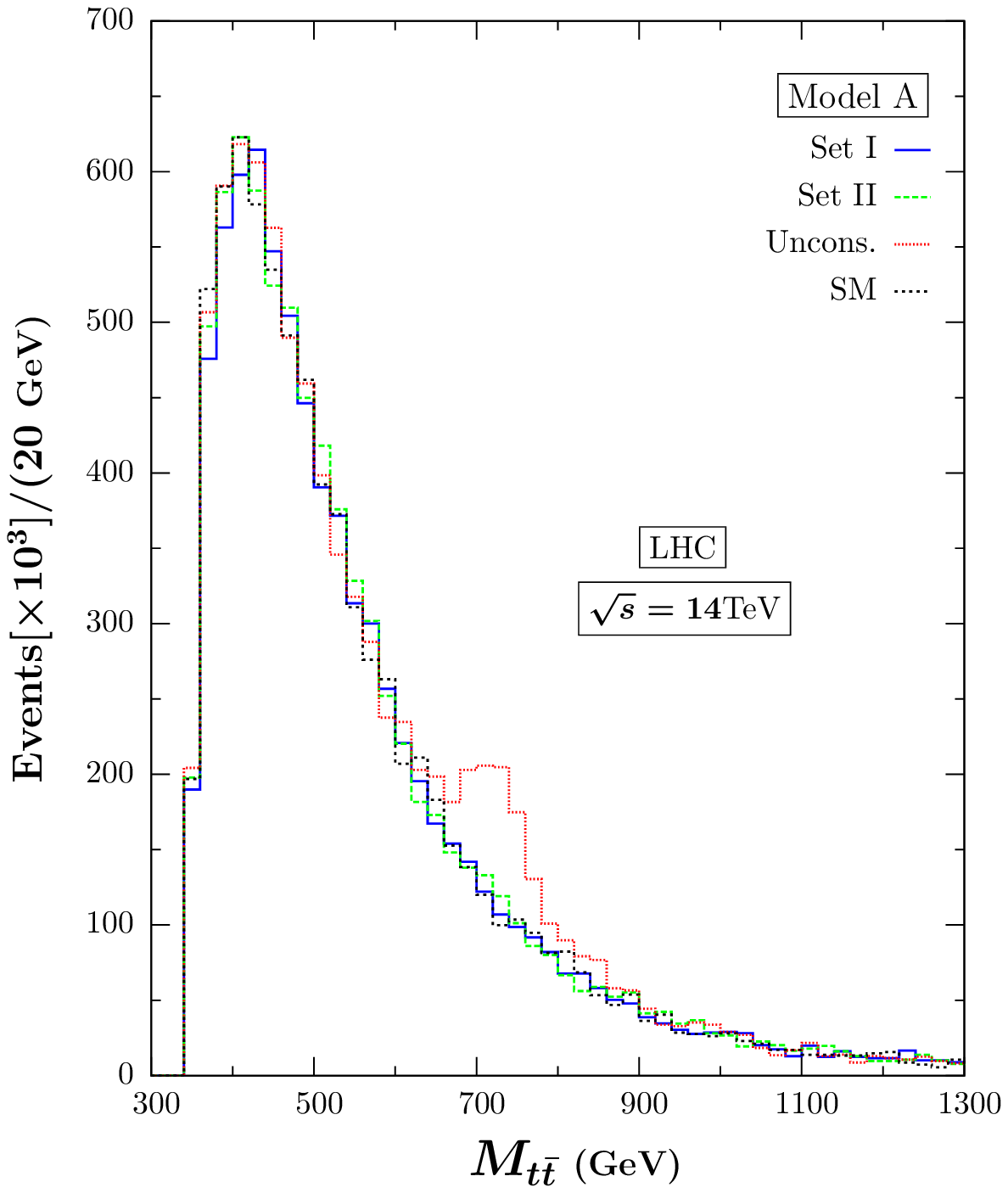} \hspace*{-0.5cm}&
\includegraphics[width=2.35in,height=2.4in]{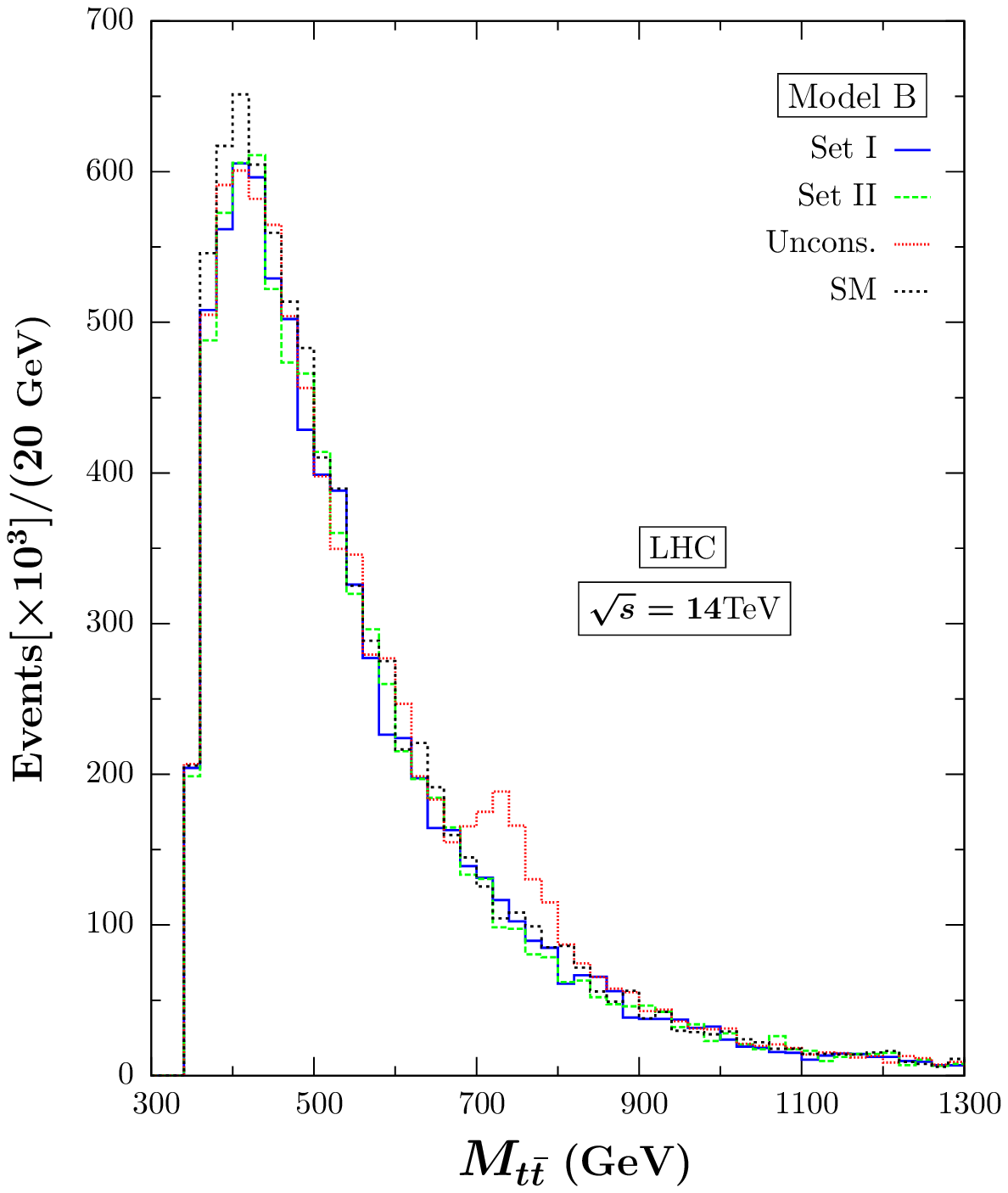}
\end{array}
$ 
\end{center}
\vskip -0.5cm
\caption{Events in the $t\bar{t}$ invariant mass distributions at LHC in Manifest LR model (left panel), Model A (middle
panel) and Model B (right panel) in comparison with the SM. Top row shows the distribution for $\sqrt{s}=7$ TeV, the bottom
row is for $\sqrt{s}=14$ TeV. Parameter sets (Set I, Set II and Unconstrained) for each model are given in the Table
\ref{params}. } 
\label{fig:minv-lhc14}
\end{figure}

We show in Fig.~\ref{fig:minv-lhc14} the number of events in the invariant mass distributions for  $t\bar{t}$ obtained after
imposing detector cuts and passing through the detector simulation, in the Manifest LR model (left panel), Model A (middle
panel) and Model B (right panel) at the LHC with $\sqrt{s}=7$ TeV (upper row) and $\sqrt{s}=14$ TeV  (lower row), where we
distinguish between Sets I, II, Unconstrained and the SM as before. These events are then used to evaluate the charge
asymmetries at the LHC. The events generated are consistent among the models studied, and show a modest bump for the
unconstrained model corresponding to the $Z_R$ resonance production. It is evident from the figure that the $M_{t\bar t}$
invariant mass distribution for all models chosen is the same, and indistinguishable from  the one in the SM. The important
distinction lies in the possible discovery of a $Z^\prime=Z_R$ boson, which in the  Manifest LR model Set I has a mass of
$1200$ GeV, as well as the ones around 730-830 GeV for the Unconstrained sets (depending on the model considered). These
appear as a resonance bump in $t\bar t$ production. For the Set I and Set II of Model A and Model B, the resonances are
heavier and out of the $M_{t \bar t}$ range presented.

We proceed to evaluate the asymmetries at the LHC. As previously mentioned, due to the $pp$ initial state, $t\bar{t}$
asymmetries at the LHC can be defined as forward and central charge asymmetries. The division of top quark rapidity $y_t$
between forward and central regions of the detector distinguishes the two asymmetries. The separation parameter $y_0$ defines
the forward $|y_t|>y_0$ and central $|y_t|<y_0$ regions of the detector. As an optimum choice of separation parameter we use
$y_0=1.5$ \cite{Hewett:2011wz}. We define the forward charge asymmetry
\begin{eqnarray}
{\cal{A}}_{F}(y_0) = \frac{N_t(y_0<|y|<2.5) - N_{\bar{t}}(y_0<|y|<2.5)}{N_t(y_0<|y|<2.5) + N_{\bar{t}}(y_0<|y|<2.5)} 
\end{eqnarray}
and the central charge asymmetry
\begin{eqnarray}
{\cal{A}}_{C}(y_0) = \frac{N_t(|y|<y_0) - N_{\bar{t}}(|y|<y_0)}{N_t(|y|<y_0) +
N_{\bar{t}}(|y|<y_0)}\ , 
\end{eqnarray}
where $N_{t(\bar t)}$ represent the number of top (anti-top) quarks with given asymmetry. To calculate the asymmetries, we
used the same procedure as in the case of the Tevatron, employing {\tt CalcHEP-Pythia-PGS} for event generation, parton
showering, jet reconstruction and detector simulation. For the analysis we used the same $lepton+jets$ topology with one
semileptonic and one hadronic top decays. We proceed by selecting single lepton events with an associated neutrino and a 
minimum 2 jets with at least one $b$-quark tagged. We imposed the following kinematical cuts for event selection at the LHC
(using the same symbols as before)
\begin{eqnarray}
 |\eta^l| < 2.5  ~~~&,&~~~ |\eta^j| < 2\ , \nonumber \\
 p_T^l > 15~{\rm GeV} ~~~&,&~~~ p_T^j > 20~{\rm GeV}\ , \nonumber \\
 \slashed{E}_T \geq 20~{\rm GeV} ~~~&,&~~~ |\eta^b| < 1\ .
\end{eqnarray}
The jets are reconstructed using a cone algorithm with $\Delta R=\sqrt{\Delta\phi^2 + \Delta\eta^2}<0.5$. Here again
$b-$jets, tagged with the loose {\tt SECVTX} algorithm, are restricted to $|\eta^b|<1$. Please note that $b$-tagging
efficiency and functions  given in {\tt PGS 4} are based on Tevatron parameters. Thus we follow the procedure  given in
\cite{Altunkaynak:2010we} to update the $b$-tagging functions according to the Eq.~(2) of \cite{Altunkaynak:2010we}. In the
LHC analysis  jet events are much more energetic due to the high center of mass energy of the collision, and thus the jet
reconstruction algorithm in {\tt PGS 4} consumes huge amount of computing time. Since the kinematical cuts are fairly relaxed
in the LHC case, we have chosen lesser amount of events ($2 \times10^{5}$) simulated for every asymmetry evaluation  with
reasonable statistical errors. After imposing all the detector cuts, asymmetries are calculated using  the $10\%$ signal
events surviving. The calculation for the LHC asymmetry in the SM as well as LR models is based on simulating events
normalized to the cross sections at NNLO level by using the standard K-factor. The results are shown in Table \ref{LHCasym}.
The asymmetries are very small, and the asymmetries in LR models can have different signs than in the SM, although
unfortunately this seems highly parameter-dependent.  At this point, these asymmetries appear  consistent (of the same
size) with the ATLAS and CMS measurements and most tend to be small and negative. To make a more definite statement, one must
wait for more precise experimental data. The LHC results are obtained over the whole rapidity parameter values, while our
results are divided into regions for better understanding of model dynamics. The experimental results have large
uncertainties, making them not yet very predictable; a higher luminosity might change that. The charge asymmetry changes
sign when measured in the forward region from the one measured in the central region of the detector in both SM and LR
models.  
\begin{table}[htb]
\begin{tabular*}{0.99\textwidth}{@{\extracolsep{\fill}} cccccc} 
\hline\hline
& & $A_{C}^{t\bar{t}}$(7 TeV) & $A_{F}^{t\bar{t}}$(7 TeV) & $A_{C}^{t\bar{t}}$(14 TeV) &
$A_{F}^{t\bar{t}}$(14 TeV)\\ 
& & ~$0<|y|<1.5$~ & ~$1.5\leq|y|< 2.5$~ & ~$0<|y|<1.5$~ & ~$1.5\leq|y|< 2.5$~ \\ \hline
\multicolumn{1}{c}{SM} &  & $-0.0024$ & $0.0157$ & $0.0011$ & $-0.0028$ \\ \hline
\multicolumn{1}{c}{\multirow{6}{*}{~LR~}} & Manifest-I & $-0.0014$ & $0.0097$ & $-0.0035$ & $0.0050$ \\ 
\multicolumn{1}{c}{}
 & Manifest-II & $0.0013$ & $-0.0091$ & $-0.0031$ & $0.0133$  \\ 
\multicolumn{1}{c}{}
 & Model A-I & $-0.0045$ & $0.0236$ & $0.0002$ & $-0.0035$  \\ 
\multicolumn{1}{c}{}
 & Model A-II & $-0.0020$ & $0.0127$ & $0.0033$ & $-0.0234$  \\ 
\multicolumn{1}{c}{}
 & Model B-I & $0.0021$ & $-0.0142$ & $-0.0002$ & $0.0003$  \\ 
\multicolumn{1}{c}{}
 & Model B-II & $-0.0001$ & $-0.0038$ & $-0.0053$ & $0.0179$  \\ \hline
\multicolumn{1}{c}{\multirow{3}{*}{$\begin{array}{c} \text{Uncons.} \\ \text{LR} \end{array}$}} & Manifest & $-0.0013$ &
$0.0063$ & $-0.0084$ & $0.0260$  \\ 
\multicolumn{1}{c}{}
 & ModelA & $-0.0117$ & $0.0650$ & $-0.0063$ & $0.0217$  \\ 
\multicolumn{1}{c}{}
 & ModelB & $-0.0087$ & $0.0469$ & $-0.0075$ & $0.0158$  \\ \cline{1-6} \hline\hline
\end{tabular*} 
\caption{Forward and Central Charge Asymmetries at LHC at signal level. Parameter sets (Set I, Set II and Unconstrained) 
for each model are given in the Table \ref{params}.}
\label{LHCasym}
\end{table}

\begin{figure}[t] 
\begin{center}
$\hspace*{-0.6cm}
\begin{array}{ccc}
\includegraphics[width=2.25in,height=2.4in]{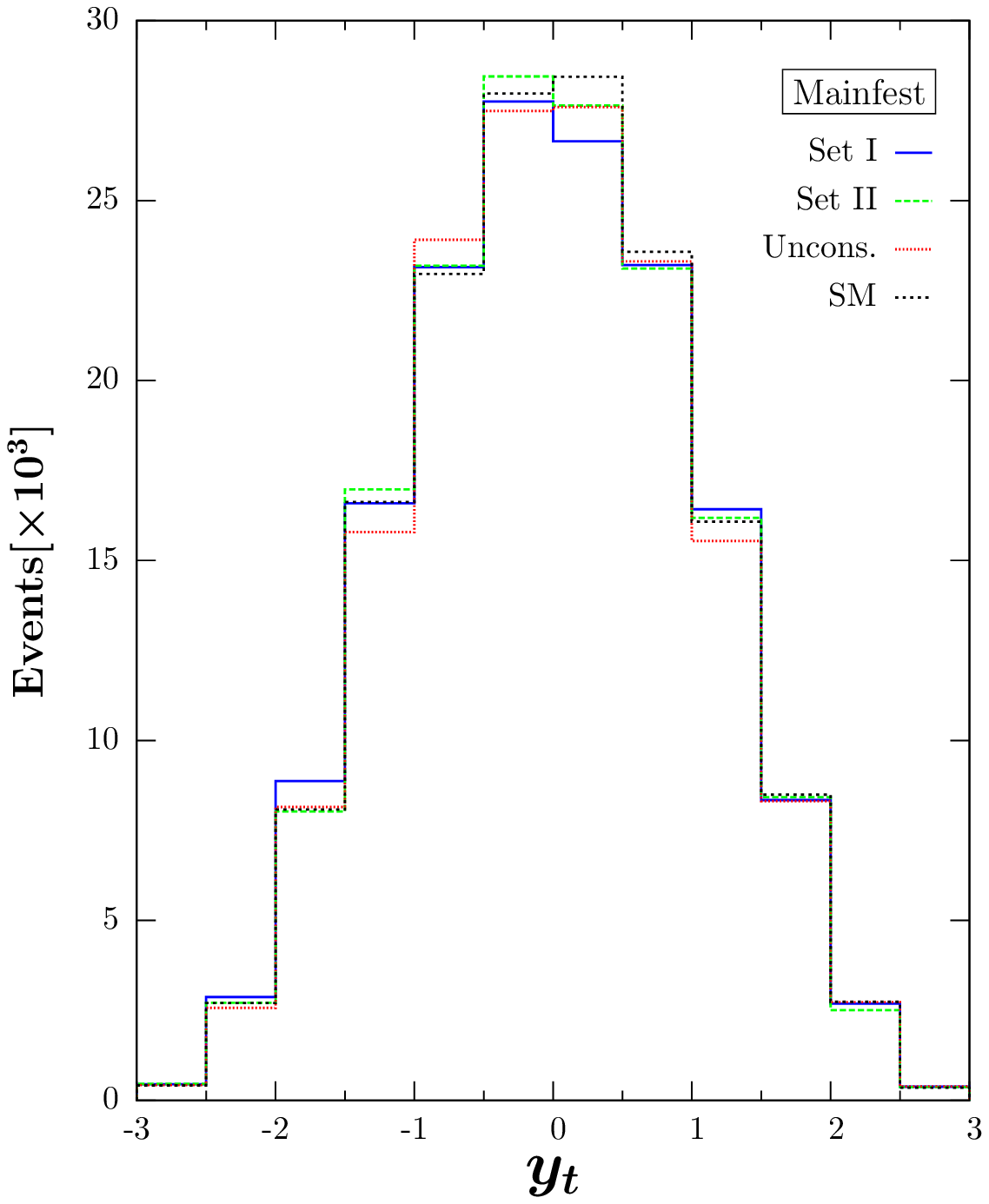} \hspace*{-0.3cm}&
\includegraphics[width=2.25in,height=2.4in]{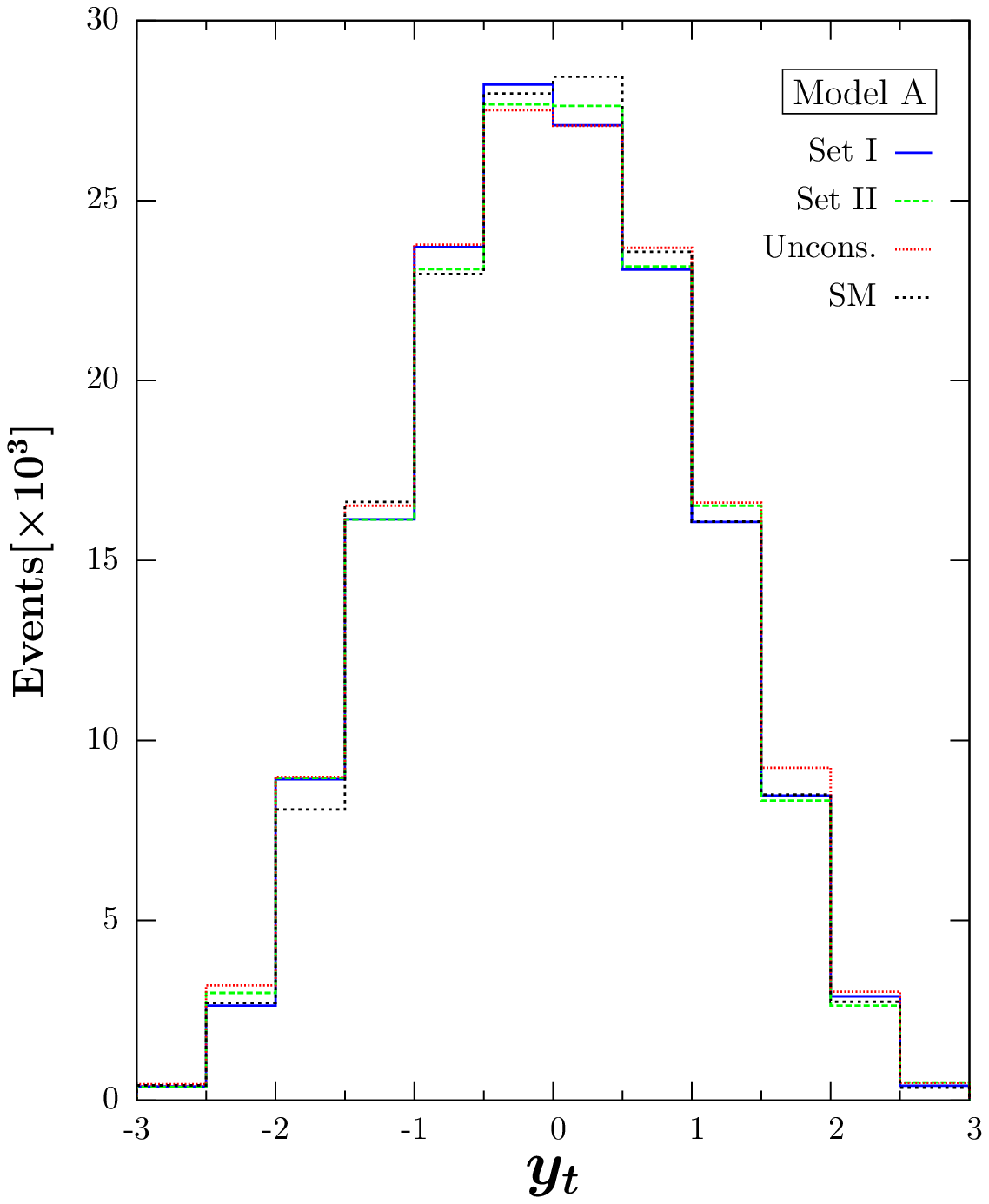} \hspace*{-0.3cm}&
\includegraphics[width=2.25in,height=2.4in]{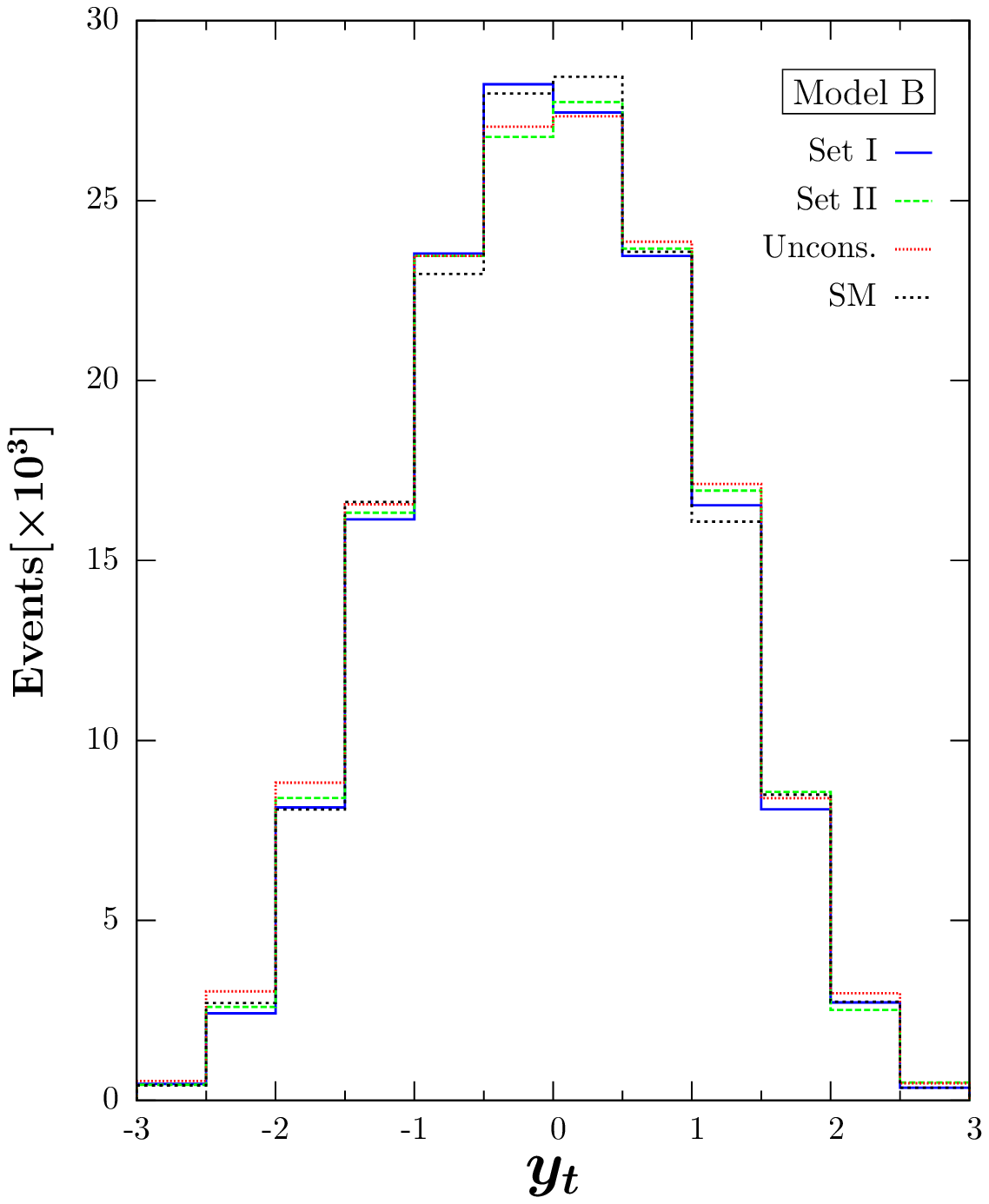} \\
\includegraphics[width=2.25in,height=2.4in]{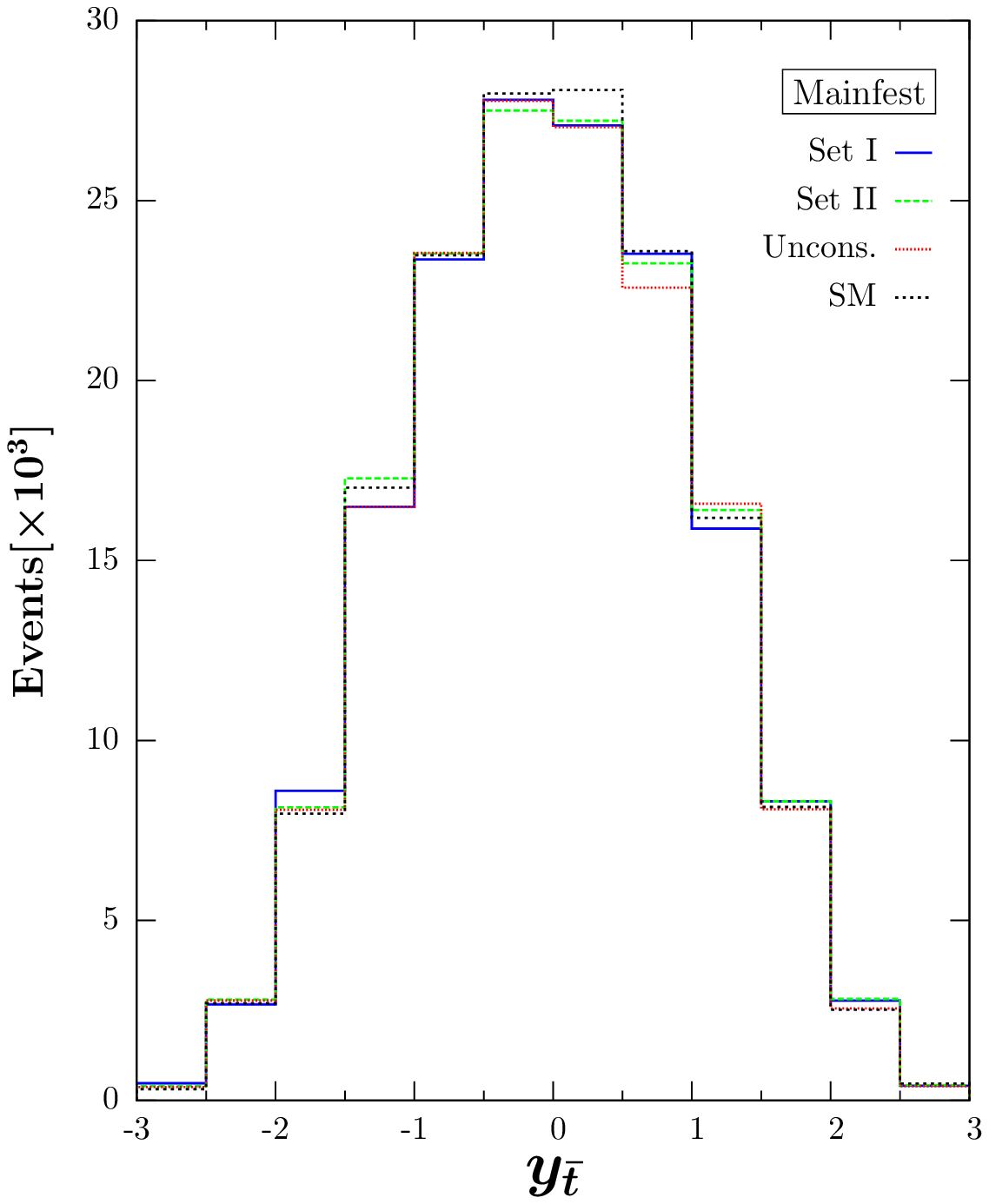} \hspace*{-0.3cm}&
\includegraphics[width=2.25in,height=2.4in]{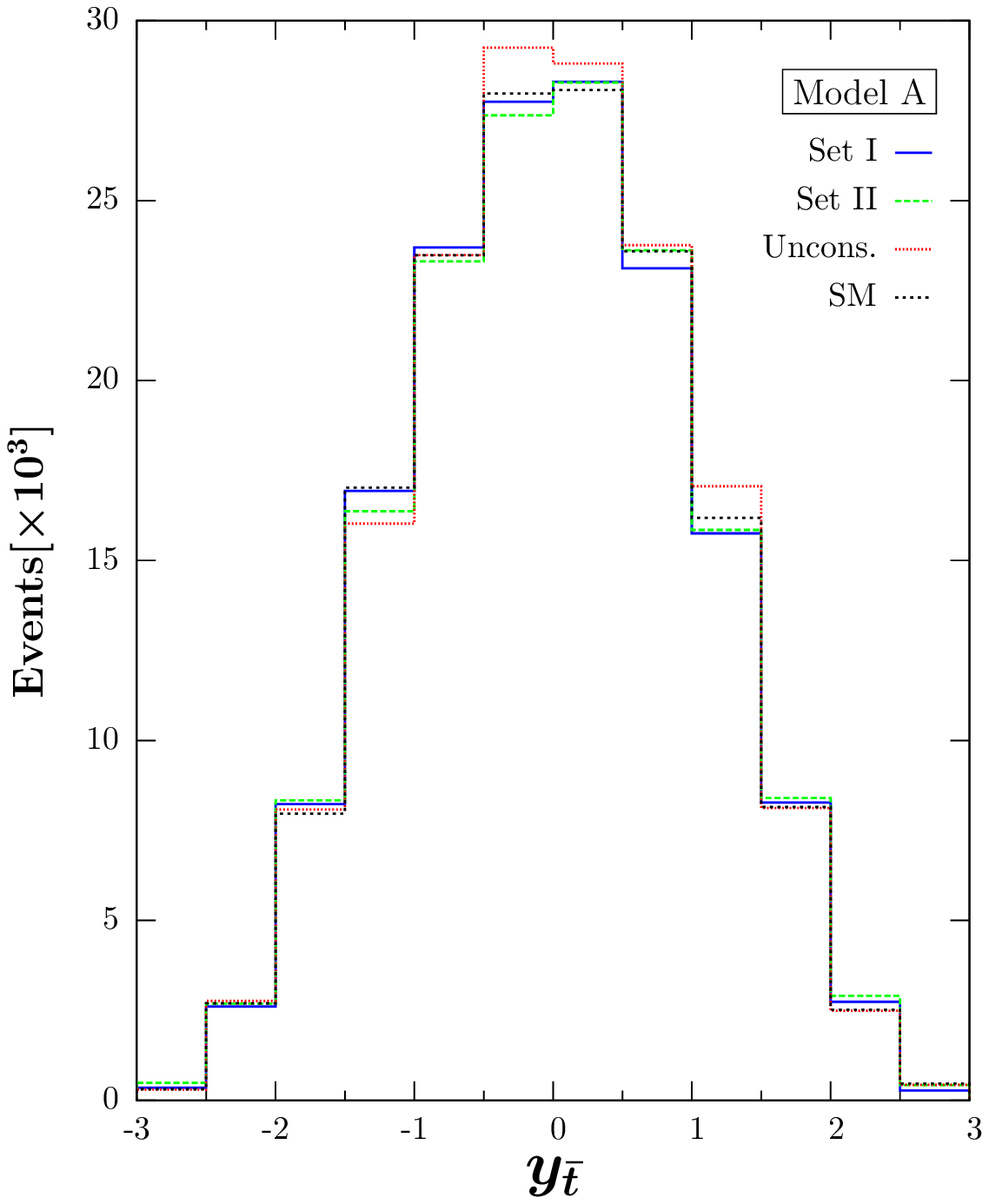} \hspace*{-0.3cm}&
\includegraphics[width=2.25in,height=2.4in]{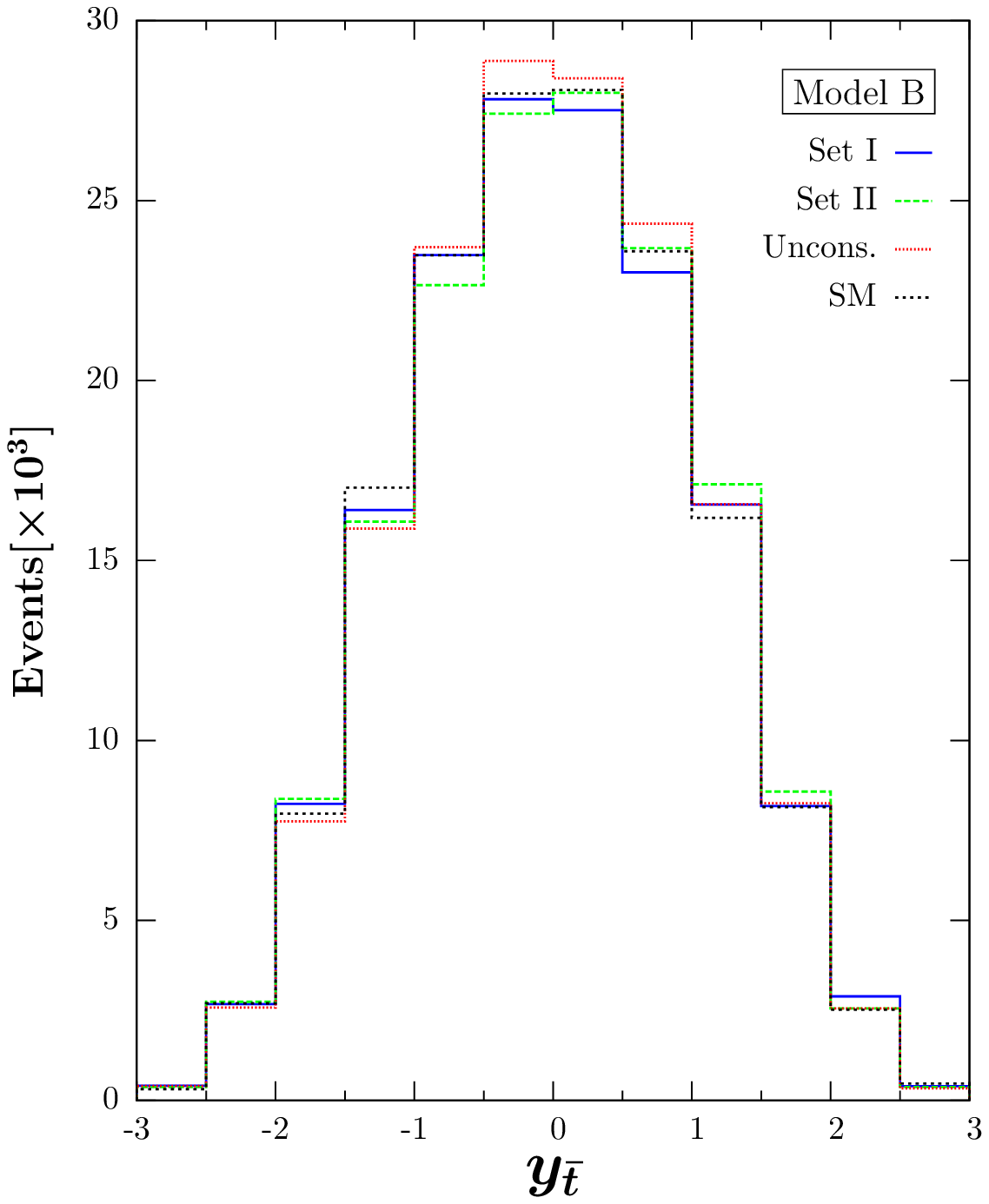}
\end{array}
$ 
\end{center}
\vskip -0.5cm
\caption{Top (upper row) and anti-top (lower row) rapidity distributions in Manifest LR model (left panel), Model A (middle
panel) and Model B (right panel) at LHC ($\sqrt{s}=7$ TeV). Parameter sets (Set I, Set II and Unconstrained) for each model
are given in the Table \ref{params}.} 
\label{fig:rapid-lhc7}
\end{figure}

\begin{figure}[t] 
\begin{center}
$\hspace*{-0.6cm}
\begin{array}{ccc}
\includegraphics[width=2.25in,height=2.4in]{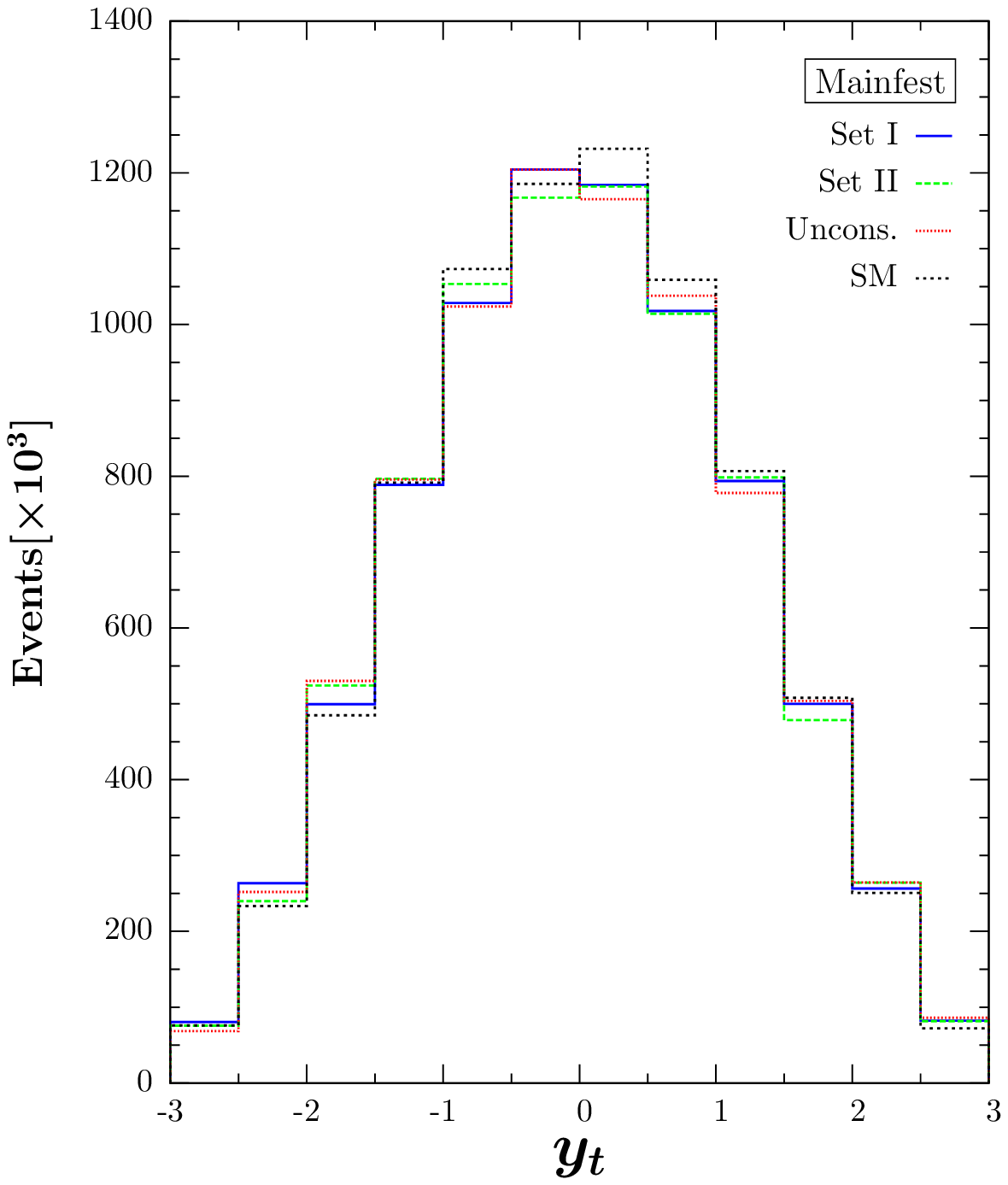} \hspace*{-0.3cm}&
\includegraphics[width=2.25in,height=2.4in]{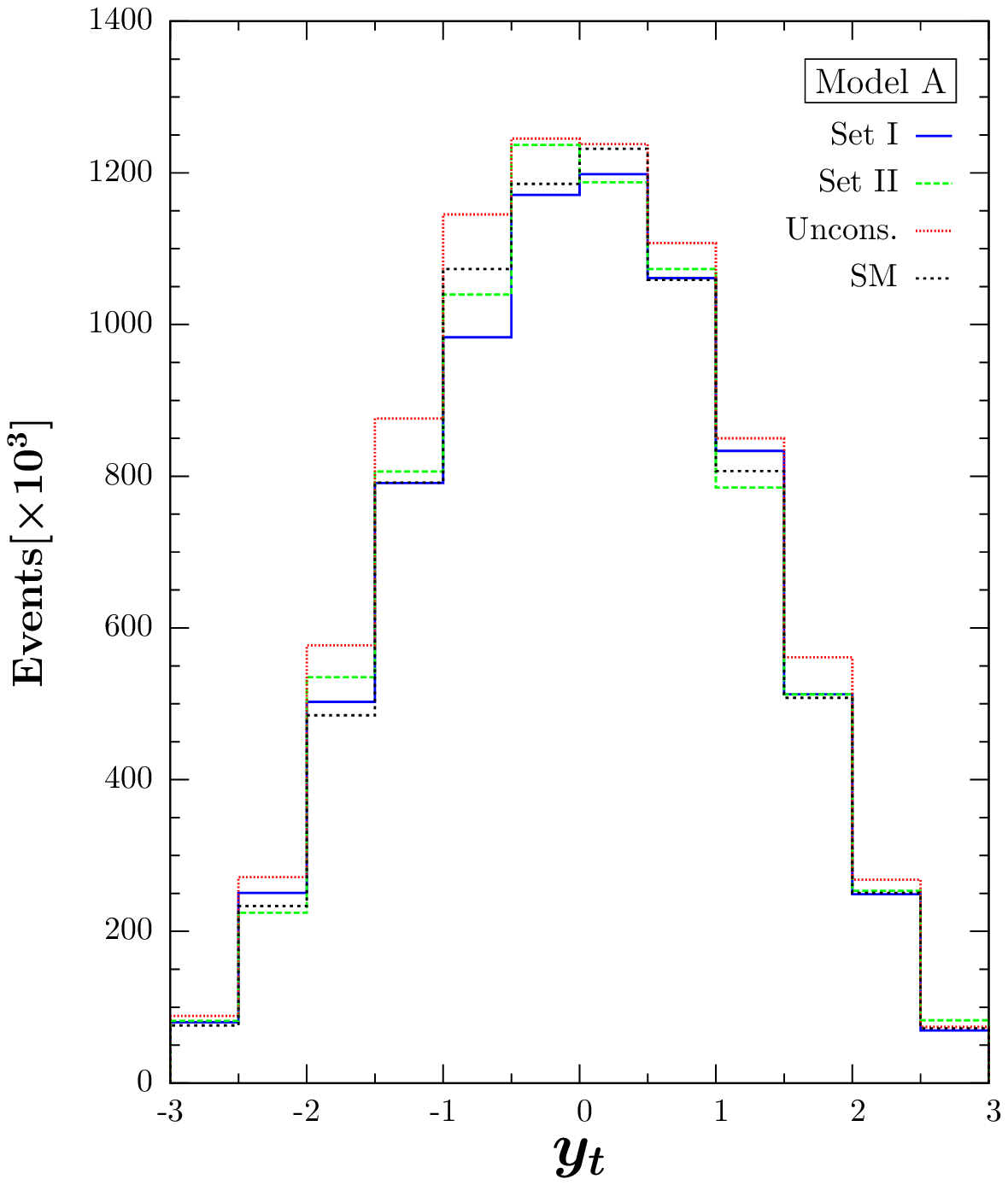} \hspace*{-0.3cm}&
\includegraphics[width=2.25in,height=2.4in]{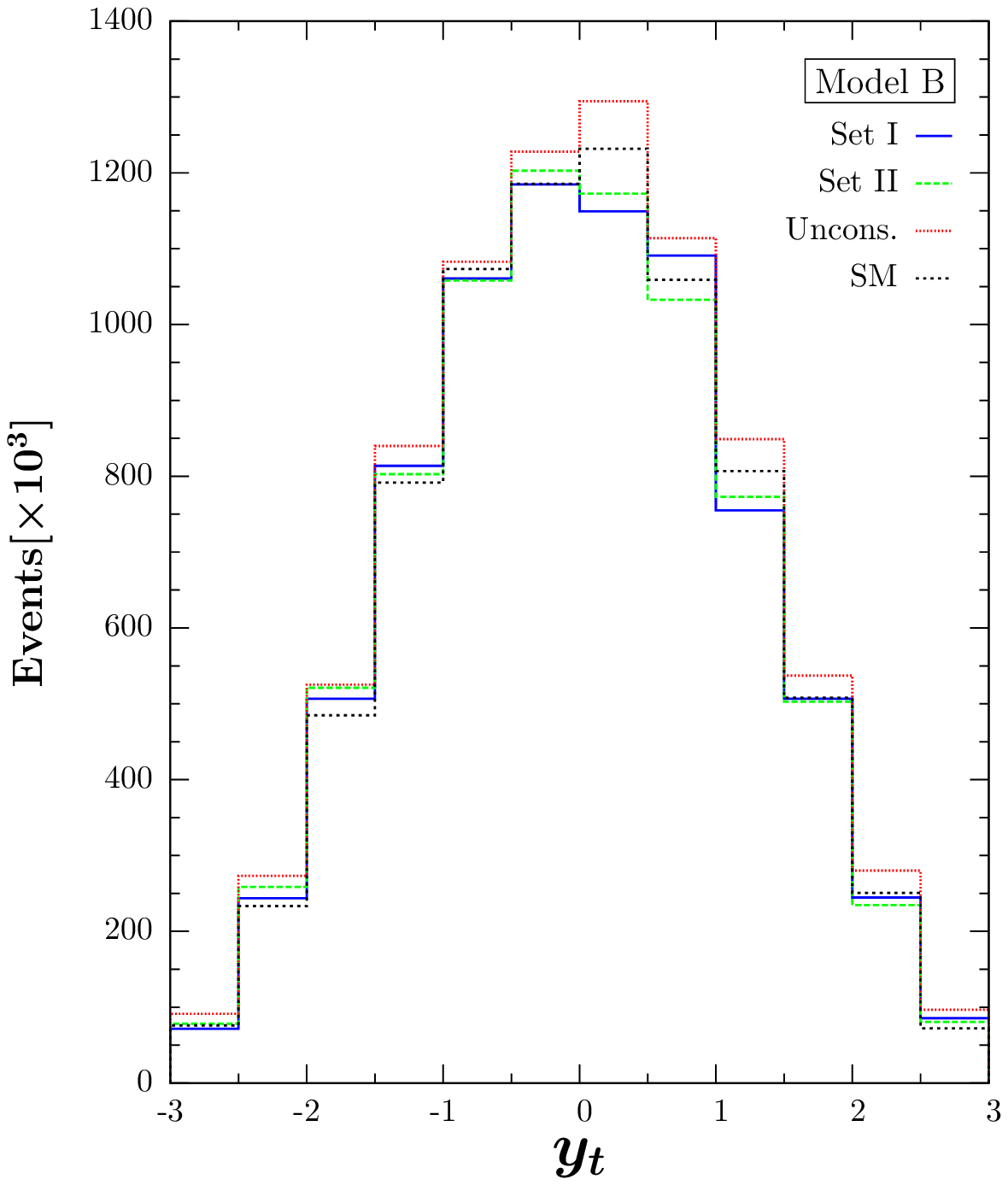} \\
\includegraphics[width=2.25in,height=2.4in]{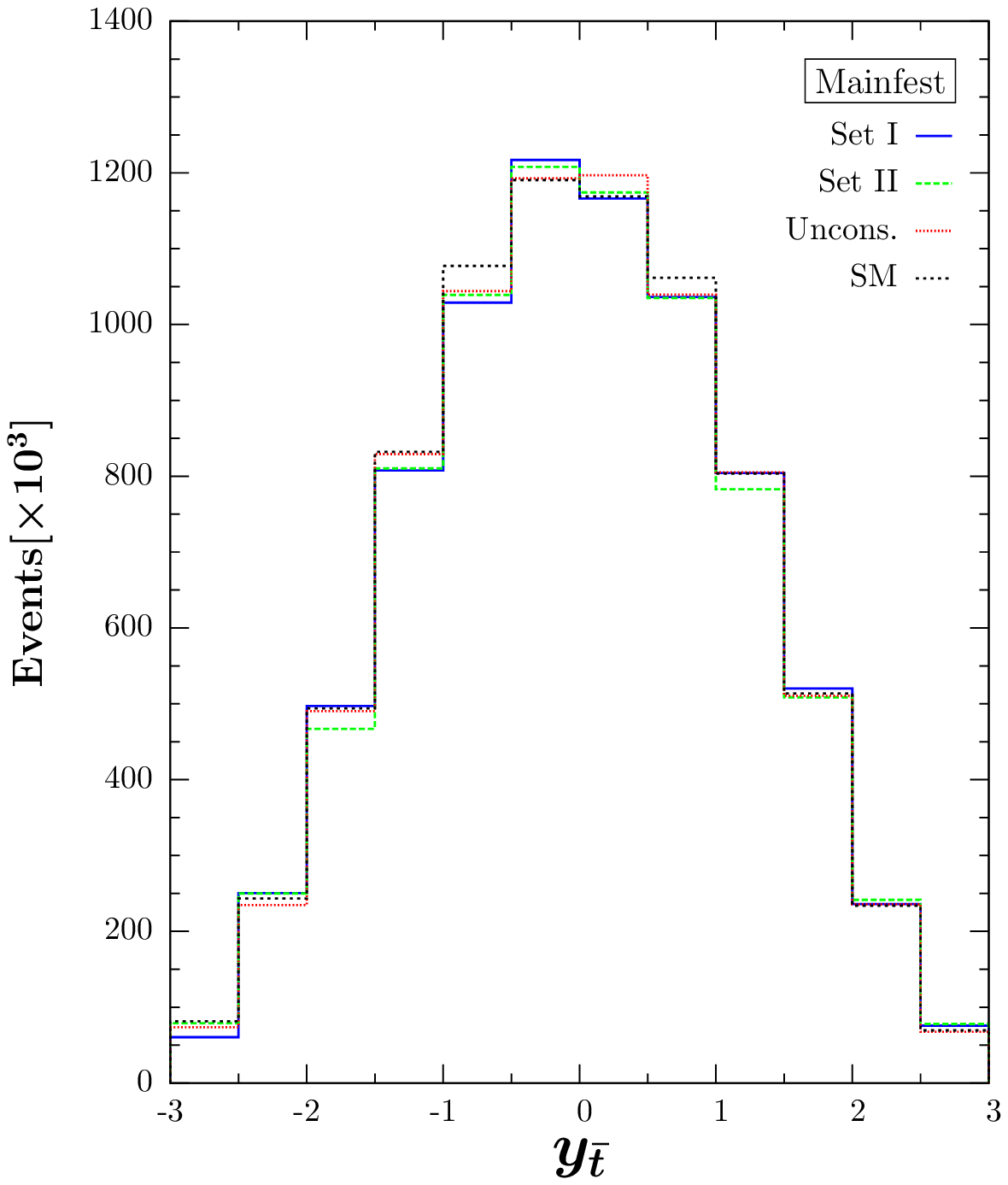} \hspace*{-0.3cm}&
\includegraphics[width=2.25in,height=2.4in]{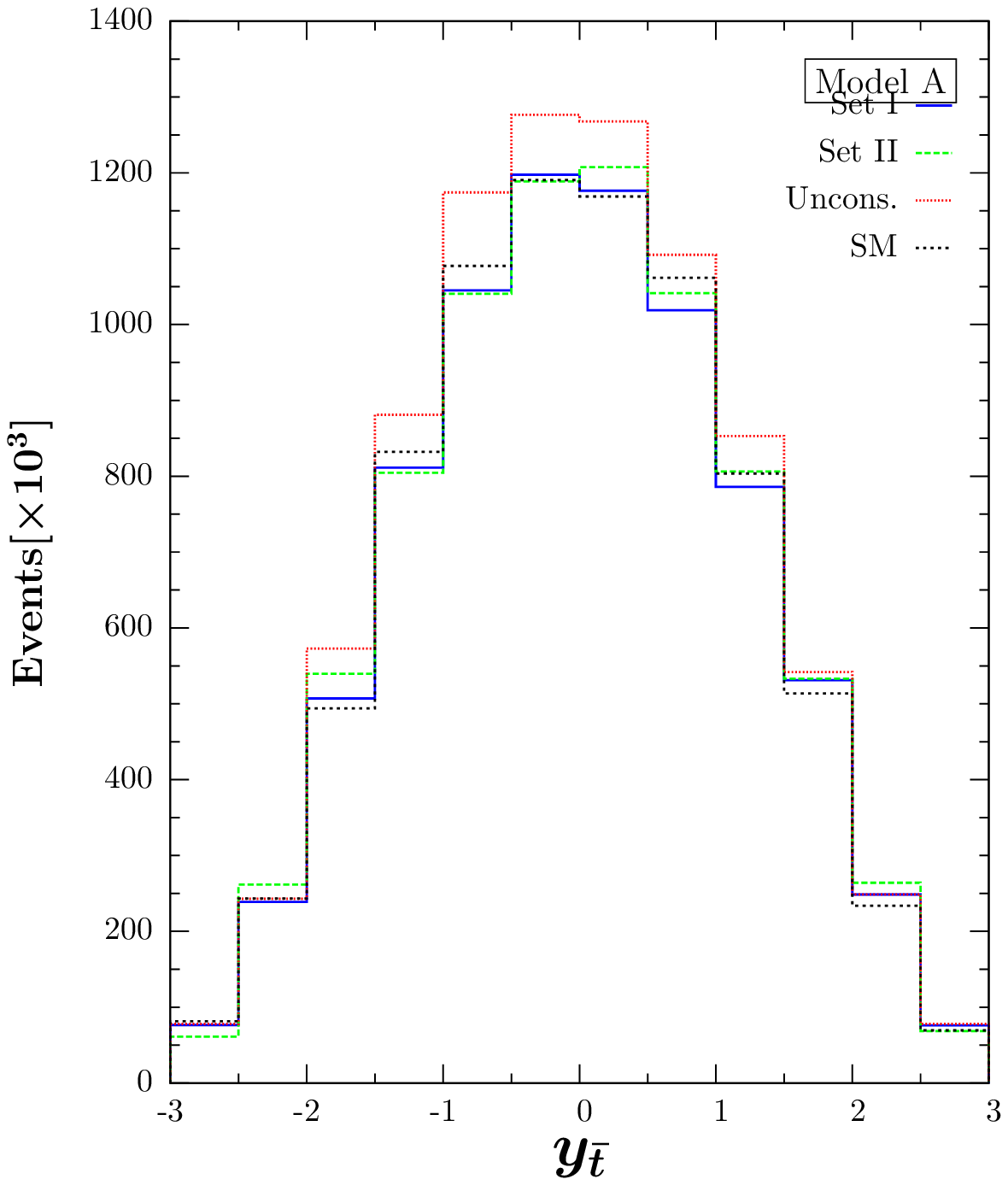} \hspace*{-0.3cm}&
\includegraphics[width=2.25in,height=2.4in]{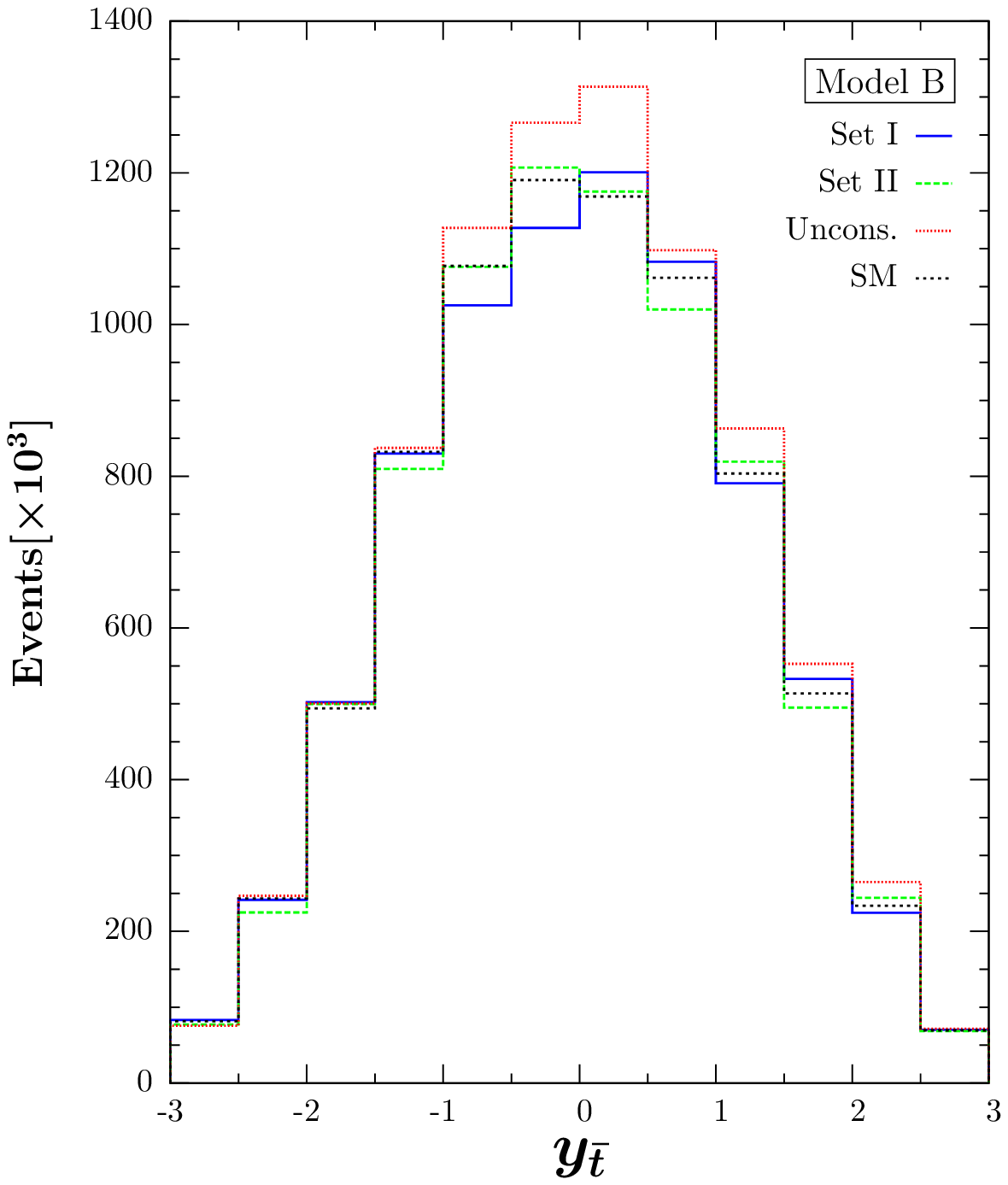}
\end{array}
$ 
\end{center}
\vskip -0.5cm
\caption{Top (upper row) and anti-top (lower row) rapidity distributions in Manifest LR model (left panel), Model A (middle
panel) and Model B (right panel) at LHC ($\sqrt{s}=14$ TeV). Parameter sets (Set I, Set II and Unconstrained) for each model
are given in the Table \ref{params}.} 
\label{fig:rapid-lhc14}
\end{figure}

In Figs.\ref{fig:rapid-lhc7} (\ref{fig:rapid-lhc14}) we show the top and anti-top rapidity distributions in LR models at the
LHC for $\sqrt{s}=7$ TeV ($\sqrt{s}=14$ TeV), in Manifest LR model (left panel), Model A (middle panel) and Model B (right
panel). Parameter sets (Set I, Set II and Unconstrained) for each model are distinguished (by blue, green and red curves).
The SM distributions are given by black curves. These figures should be compared to Fig.~\ref{fig:rapid-tev} from the
Tevatron section. By comparison,  the LHC asymmetries are even more dominated by events at, or near zero charge asymmetry for
both top and anti-top quarks and do not show measurable deviations in LR models. Thus a significant charge asymmetry for top
or anti-top quarks at the LHC would be indicative of BSM scenarios other than left-right models-- so far, this does not
appear to be the case. It may be difficult to use the charge asymmetry to distinguish between various models, even those
which predict large asymmetries at the Tevatron, as a comprehensive analysis of their predictions at the LHC shows that they
seem to be small, though some models may differ when evaluated at high invariant masses, which are especially sensitive to
the $q {\bar q}$ contribution \cite{AguilarSaavedra:2011ug}.

\section{Conclusions}
\label {sec:conclusion}

The observation of a large forward-backward asymmetry in $t {\bar t}$ production at the Tevatron offers tantalizing signals
of physics beyond the Standard Model. For large rapidities and large invariant $t {\bar t}$ mass distributions, the
measurements deviate by 3$\sigma$ or more from the SM expectations. This seems to indicate that the phenomenology of the top
quark, which has a mass of the order of electroweak symmetry breaking, may offer a window into new much anticipated BSM.
Several models have been produced specifically to deal with the measurements. Though instructive, they seems like a band-aid
solution. In addition, recent investigation of whether the increase in the asymmetry at large invariant mass $M_{t {\bar t}}$
can be accounted for by a tree-level scalar exchange indicates that the range of models who remain consistent with other
top-related measurements, flavor violation constraints, electroweak precision measurements and collider data, is far more
restricted than initially thought. There are at present other measurements  which indicate deviations from the SM, which are
not explained by most of the {\it ad-hoc} models which provide a fix for the forward-backward asymmetry. 

One can then ask, what about the BSM scenarios favored on theoretical grounds, and already analyzed  and subjected to
relevant phenomenological and experimental tests. In this work, we analyze the left-right model, in fact a general version of
this model, where left and right coupling constants are not equal, and the quark mixing matrices in the left and right
sectors are unrelated. The model is subjected to constraints coming from meson mixing ($K^0-{\bar K}^0$, $B_d^0-{\bar B}_d^0$
and $B_s^0-{\bar B}_s^0$) and $b \to s \gamma$. The production of $W_R$ has been previously studied in this model and limits
on the masses, coupling constants and right-handed quark mixing have been included. It is worthwhile to ask whether such a
model can explain the deviation of the predicted asymmetry from the observed one  at the Tevatron. The LR model has the
features desired for a resolution: a $W_R$ in the $t$-channel which can be responsible for the asymmetry, and a heavier $Z_R$
in the $s$-channel, which may affect the observed cross section. 
 
Our analysis shows that, if the cross section agrees with the SM model one, as confirmed by the CDF data, the model is not
able to generate sufficient asymmetry at the Tevatron to explain the observed discrepancy. We should add that this result
survives variations in coupling constants, boson masses and right-handed CKM mass mixing parameters in the allowed parameter
space determined by low-energy data. Relaxing these constraints would definitely yield bigger asymmetries  and would provide
large enough asymmetries to agree with the Tevatron data, as the Unconstrained version of LR models shows. This model is thus
unlike models which explain the asymmetry through exchange of a light $W'$ in the $t$-channel, coupling with a large coupling
to only the $t-d$ quark sector, and which requires additional fermions for anomaly cancellation.
 
We analyze the $t {\bar t}$ cross section and asymmetries at the LHC. The cross section agrees with the one predicted by SM
and measured at $\sqrt{s}=7$ TeV. One would expect to see the $Z_R$ resonance for increased CM energy: so far, the
indications are negative, pushing the $Z^\prime$ mass into the TeV range (although the precise values depend on the model and
parameters chosen). It is also likely that the LHC, looking for top jet resonances,  would either validate or  rule out at
$>3\sigma$ level any extra $Z'$ or $W'$ models which can reproduce the Tevatron asymmetry. The left-right models predict a
negligible charge asymmetry (the relevant defined parameter at the LHC), in either forward or central regions,  at both
$\sqrt{s}=7$ and $14$ TeV. The predictions for the asymmetry are not always well-defined in sign, but the LR models are
consistent with the SM predictions and so far, with the experimental results form ATLAS and CMS. The forward and central
charge asymmetry have opposite signs. The arbitrariness in sign is unfortunate as it was shown that a definite positive
(central-value) charge asymmetry at the LHC would strengthen the Tevatron results, while a definite negative (central-value)
asymmetry would be unexpected and its explanation conflict with models that pass the Tevatron requirements
\cite{AguilarSaavedra:2011ug}. One can draw two conclusions. One is that while the LR models predictions for the cross
sections at the Tevatron and LHC {\it and} the asymmetry at LHC agree with the experimental data, these models cannot provide
an explanation for the observed Tevatron forward-backward asymmetry. We can ascertain this with confidence, as it is valid
for a large region of the parameter space and valid  independent of whether we chose Manifest, Model A or Model B. The
questions still remain: are the Tevatron and LHC results inconsistent with each other (this will become clear with more
precise LHC data), and what is the origin of the large forward-backward  asymmetry. The second conclusion is  that, while
predictions for  charge and forward-backward asymmetries are important in comparing models to experimental data, they not
good indicators of left-right models  because they are very small. A more promising alternatives would be to search for
$W_R$ bosons, predicted to be lighter than $Z_R$; measurements of top quark polarization which could indicate right-handed
physics; and measuring  left-right, rather than forward-backward, asymmetries. These tests are beyond the scope of this work
and will be presented elsewhere. 
 
There is however another issue that arises. Except for the {\it ad-hoc} models (some of which are already ruled out by a more
careful analysis), it appears likely that none of the better-known BSM scenarios can produce  large  forward-backward
asymmetries. Should negative asymmetries survive at LHC, consistency with Tevatron measurements would be 
challenging and demonstrate that top quark physics  has subtleties not fully yet understood. Should  asymmetries at the
LHC  be found to be small and positive,  the challenge would be in how to understand their enhancement in $p {\bar p}$ but
not $pp$ (within normal expectations of symmetries in $pp$ initial states). But before measurements, one must know what
results to expect from  established BSM scenarios.  As many such scenarios are plagued by uncertainties due to a large
parameter space, a clear result is important, as it would restrict BSM possibilities.

\section*{Acknowledgements}
M.F. and A.H. would like to thank   NSERC of Canada for partial financial support under grant number SAP105354. 
 

\end{document}